\documentclass[twocolumn]{emulateapj}
\usepackage{color}

\usepackage[usenames,dvipsnames,svgnames,table]{xcolor}
\usepackage[colorlinks=true,
           linkcolor=blue,
           urlcolor=blue,
           citecolor=blue]{hyperref}

\begin{document}

\title{FORMATION AND ERUPTION OF A SMALL FLUX ROPE IN THE CHROMOSPHERE OBSERVED BY NST, IRIS, AND SDO}

\author{PANKAJ KUMAR\altaffilmark{1}, VASYL YURCHYSHYN\altaffilmark{2,1}, HAIMIN WANG\altaffilmark{3,2}, KYUNG-SUK CHO\altaffilmark{1,4}}
\affil{$^1$Korea Astronomy and Space Science Institute (KASI), Daejeon, 305-348, Republic of Korea}
\affil{$^2$Big Bear Solar Observatory, New Jersey Institute of Technology, Big Bear City, CA 92314, USA}
\affil{$^3$Space Weather Research Laboratory, New Jersey Institute of Technology, University Heights, Newark, NJ 07102-1982, USA}
\affil{$^4$ University of Science and Technology, Daejeon 305-348, Republic of Korea}
\email{pankaj@kasi.re.kr}

\begin{abstract}
Using high-resolution images from 1.6 m {\it New Solar Telescope} (NST) at Big Bear Solar Observatory (BBSO), we report the direct evidence of chromospheric reconnection at the polarity inversion line (PIL) between two small opposite polarity sunspots. Small jet-like structures (with velocities of $\sim$20-55 km s$^{-1}$) were observed at the reconnection site before the onset of the first M1.0 flare. The slow rise of untwisting jets was followed by the onset of cool plasma inflow ($\sim$10 km s$^{-1}$) at the reconnection site, causing the onset of a two-ribbon flare. The reconnection between two sheared J-shaped cool H$\alpha$ loops causes the formation of a small twisted flux rope (S shaped) in the chromosphere. In addition, Helioseismic and Magnetic Imager (HMI) magnetograms show the flux cancellation (both positive and negative) during the first M1.0 flare. The emergence of negative flux and cancellation of positive flux (with shear flows) continue until the successful eruption of the flux rope. The newly formed chromospheric flux rope becomes unstable and rises slowly with the speed of $\sim$108 km s$^{-1}$ during a second C8.5 flare that occurred after $\sim$3 hours of the first M1.0 flare. The flux rope was destroyed by repeated magnetic reconnection induced by its interaction with the ambient field (fan-spine toplology) and looks like an untwisting surge ($\sim$170 km s$^{-1}$) in the coronal images recorded by Solar Dynamic Observatory/{\it Atmospheric Imaging Assembly} (SDO/AIA). These observations suggest the formation of a chromospheric flux rope (by magnetic reconnection associated with flux cancellation) during the first M1.0 flare and its subsequent eruption/disruption during the second C8.5 flare.
   
 \end{abstract}

\section{INTRODUCTION}
A magnetic flux rope consists of twisted helical field lines that are wrapped around its central axis. Flux ropes are considered an important part of a coronal mass ejection (CME), which plays a crucial role in the propagation of a CME in the interplanetary medium. These flux ropes are identified as magnetic clouds at 1 AU, and produce severe geomagnetic storms by reconnecting with the earth's magnetosphere \citep{burlaga1982,yurchyshyn2001,manoharan2010,kumar2011,marubashi2012,cho2013}. Therefore, the study of flux ropes is crucial from space-weather prospective.

Understanding the formation mechanisms of a magnetic flux rope on the sun is very important for many CME initiation models. There is a long lasting debate on the formation mechanism of a magnetic flux rope, i.e., Are flux ropes formed during the magnetic reconnection or they are emerging below the photosphere? The CME initiation models are classified into two categories. Many CME models such as emerging flux \citep{chen2000}, kink or torus instabilities \citep{fan2004,torok2005,kliem2006} consider a pre-existing flux rope that emerges from below the photosphere, whereas in other models (i.e., tether cutting \citep{moore2001}, magnetic breakout\citep{antiochos1999,karpen2012}) the twisted flux rope is formed during magnetic reconnection in the corona. Topological models proposed by \citet{gosling1995} and \citet{longcope2007} also show the formation of a helical flux rope by successive magnetic reconnection of the arcade loops in a two ribbon flare. 

Filaments or prominences contain cool plasma and are frequently observed in the chromosphere along the neutral line. According to the flux rope model, cool plasma of filaments or prominences is supported in the dips of a helical flux rope by magnetic tension \citep{van1989, priest1989}. The formation of a filament may be due to emergence of U-shaped loops joining already emerged segment of a flux rope below the photosphere \citep{rust1994}.  Alternatively, successive flux cancellation at the PIL driven by photospheric shear motion can generate a helical flux rope from the coronal sheared arcades \citep{van1989}.
Flux emergence simulations \citep{fan2001, archontis2008} reveal that the emergence of flux rope stops when its magnetic axis reaches the photosphere and produces sheared arcades.
The flux cancellation process may be important at the locations where the emerging flux rope is unable to cross the photosphere, therefore, forming of flux rope in situ along the PIL \citep{amari2003,amari2010,aulanier2010,amari2014}. 

There is a number of observational evidence that support the flux rope model of a prominence/filament. For example, when prominences are viewed on the limb along its axis, a tunnel like elongated (dark) structure is frequently observed in the EUV, soft X-ray and white light images. These low density dark structures are known as coronal cavities \citep{gibson2006a,gibson2006b}. Observations from COMP (Coronal Multi-channel Polarimeter) reveal that these coronal cavities are consistent with the flux rope model \citep{bak2013}. In addition, the three part structure of a CME consists of a bright frontal loop, a dark cavity and a bright core (filament/prominence material) \citep{hund1999}. The circular features observed in CMEs in the coronagraph images are basically the evidences of a flux rope \citep{vourlidas2013,vourlidas2014}.  

Using Hinode SOT observations, \citet{okamoto2008} reported the evidences of the emergence of a helical flux rope from below the photosphere into the corona (along the PIL) under a preexisting prominence. \citet{kumar2013r} observed the emergence of a bipole with rotation in opposite direction and simultaneous appearance of a twisted flux rope in the EUV images, suggesting its emergence below the photosphere. Using Hinode XRT (X-ray telescope) images and photopsheric magnetograms \citet{green2009} reported the convergence and cancellation of magnetic flux at the PIL leading to the formation of a flux rope before the eruption. Simultaneous appearance of a helical flux rope in the hot (XRT, T$\sim$10 MK) and cool EUV (171, 304 \AA) channels, supports the flux rope model that the filament cool plasma is supported by a helical flux rope \citep{kumar2011a}.   
At present, there are many observational reports on the flux rope formation during the eruption observed only in the hot AIA channels (131 and 94 \AA) \citep{cheng2011,zhang2012}. \citet{kumar2014} observed the formation of a twisted flux rope during magnetic reconnection above a kink unstable  small filament. Reconnection above the filament destroes it and a newly formed twisted flux rope successfully erupted to launch a high speed CME. Note that the observational reports discussed above are related to the flux rope formation in the corona.
Recently, Using high-resolution NST data \citet{wang2015} reported the detailed structure and evolution of a confined S-shaped flux rope in the chromosphere.
 The direct observation of the formation of a twisted flux rope in the chromosphere is very important and should be investigated in more details.

In this paper, we mainly focus on the formation and eruption of a small flux rope in the chromosphere within active region (AR) NOAA 12087 on 12-13 June 2014. Understanding the formation mechanism of a small flux rope (and their role in the flare trigger) may help in understanding the formation of large flux ropes involved in huge CMEs. At present, these small flux rope can be resolved with high-resolution observations made by NST and IRIS instruments. We utilized SDO/AIA and RHESSI hard X-ray images to investigate the coronal magnetic field configuration and particle acceleration site during the flares. 
In Section~2, we present the observations and results. In the last section, we summarize and discuss the results.

\begin{figure*}
\centering{
\includegraphics[width=8cm]{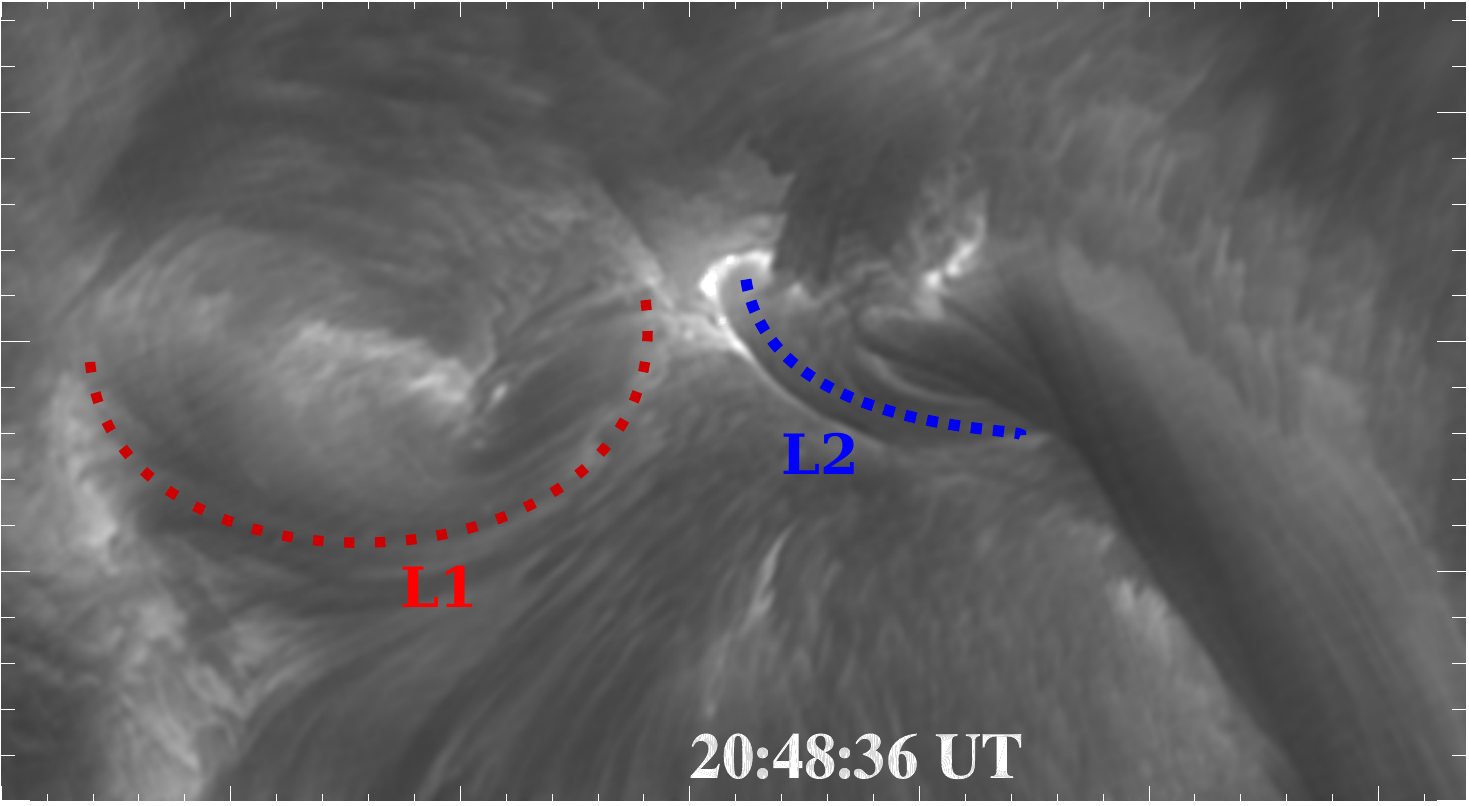}
\includegraphics[width=8cm]{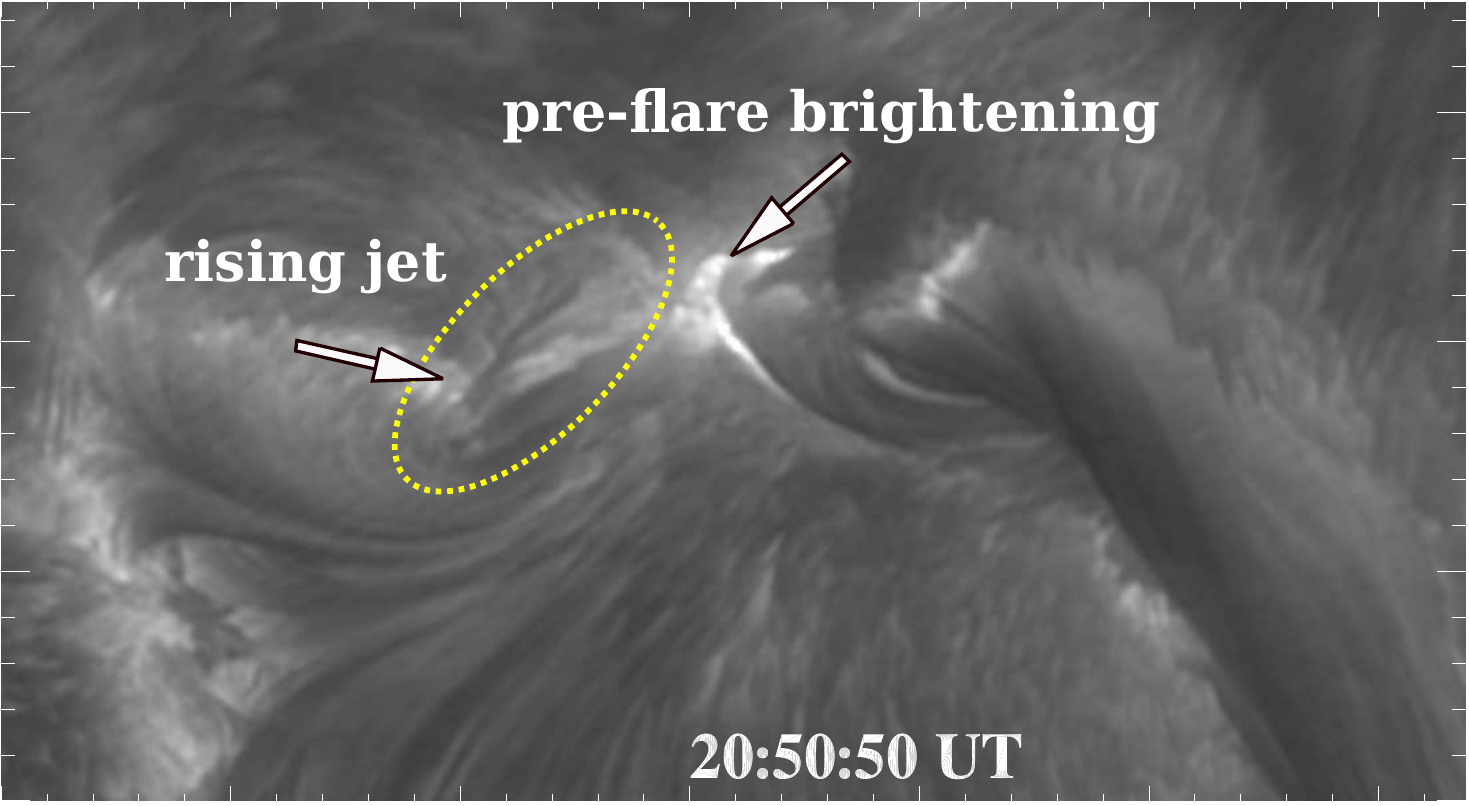}

\includegraphics[width=8cm]{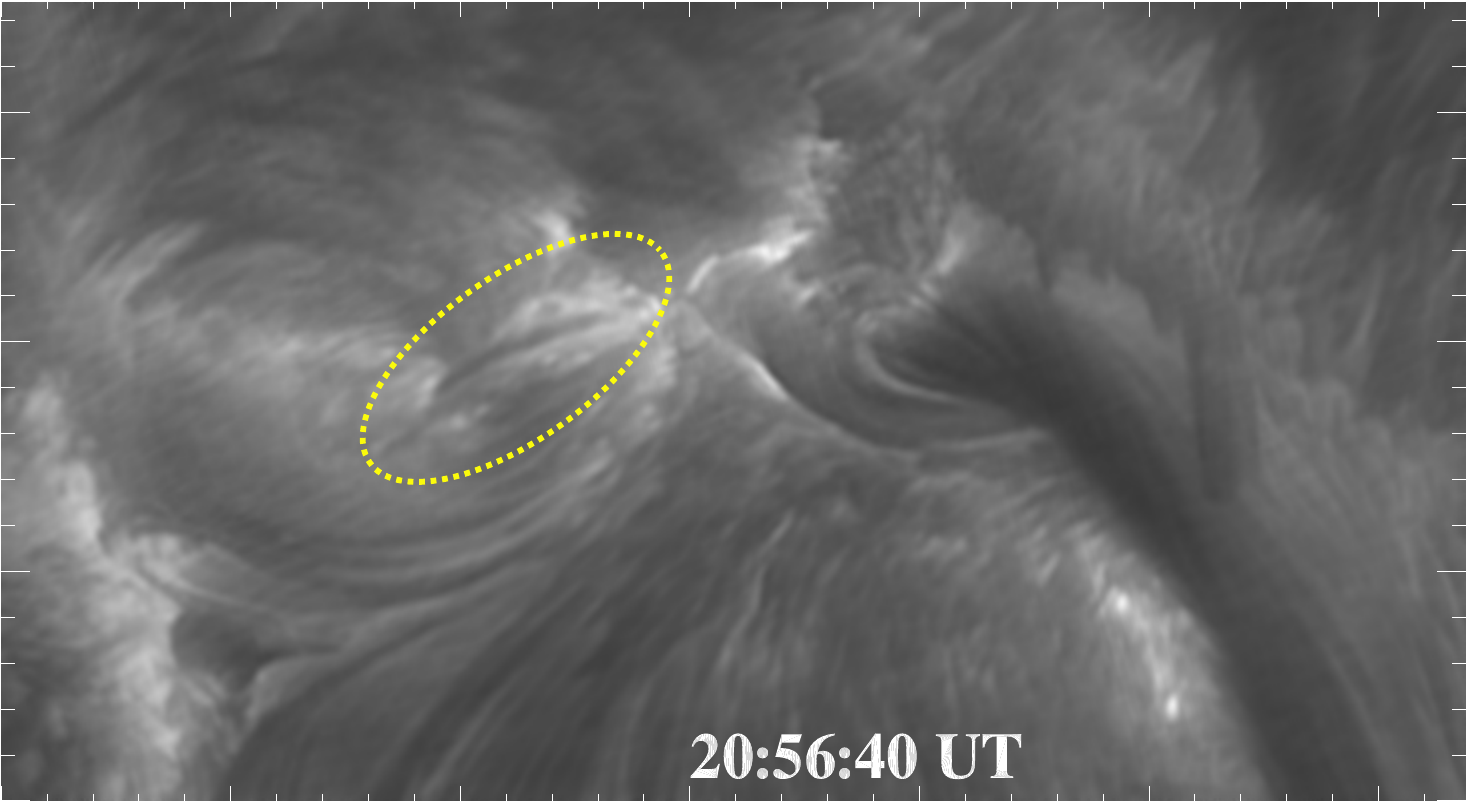}
\includegraphics[width=8cm]{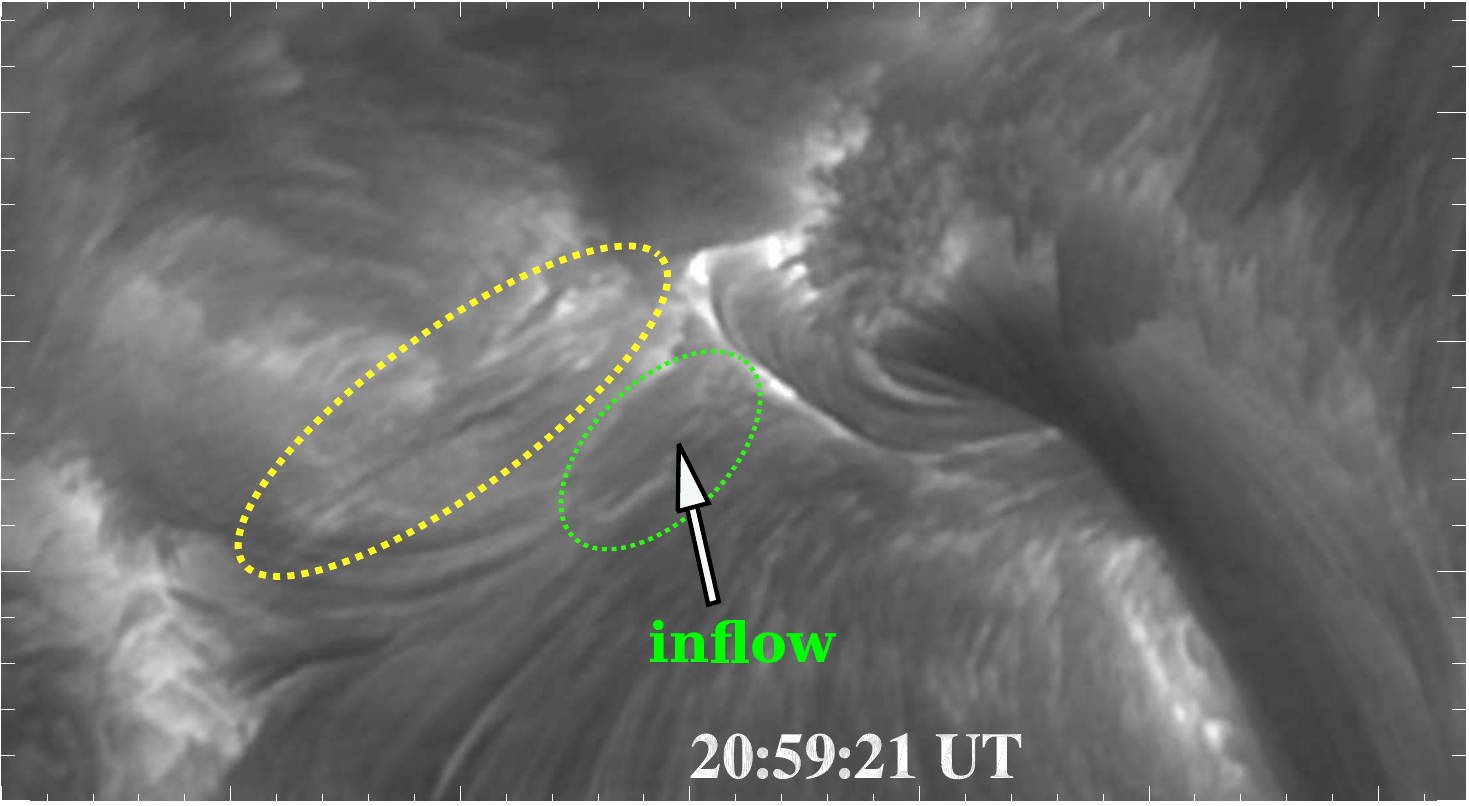}

\includegraphics[width=8cm]{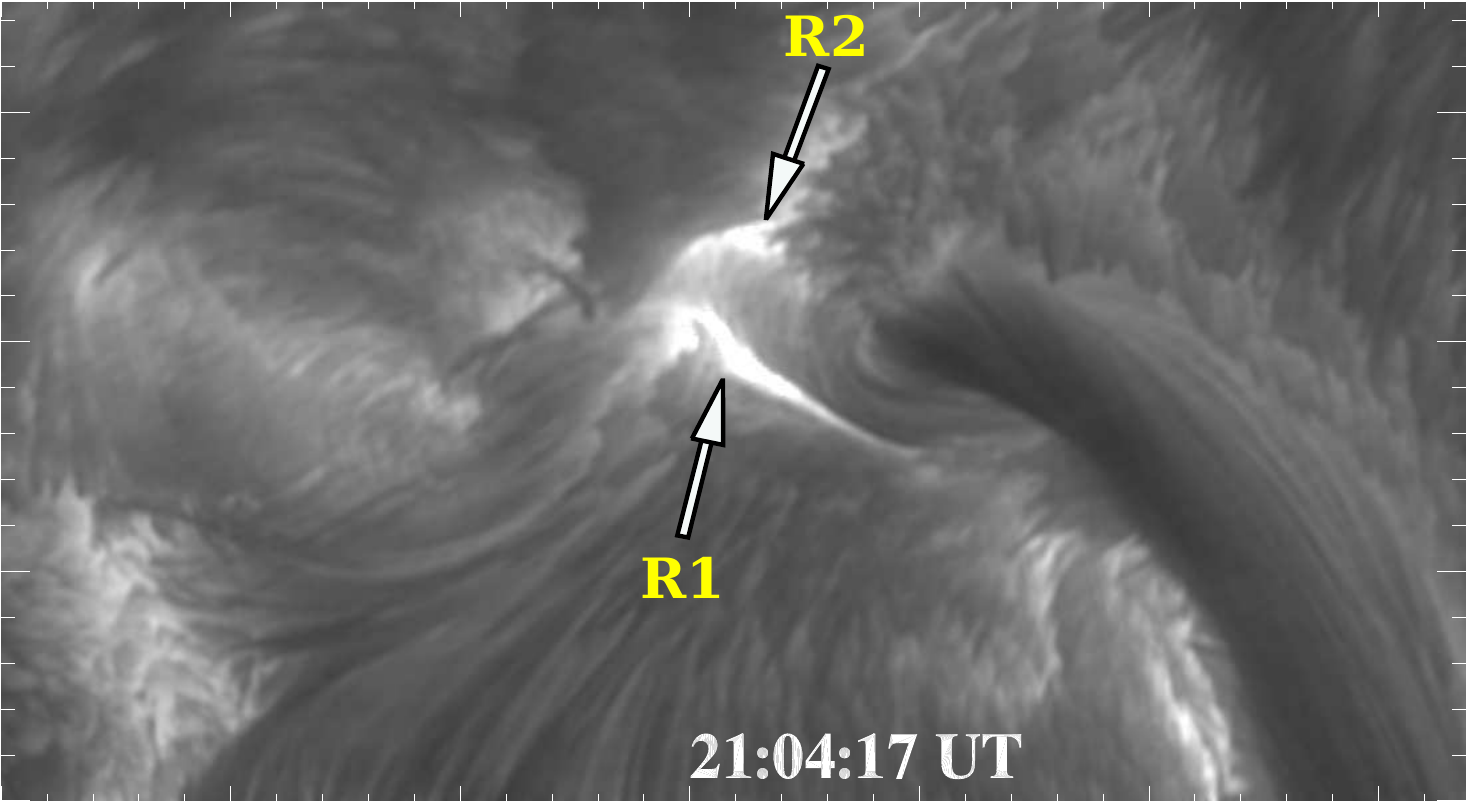}
\includegraphics[width=8cm]{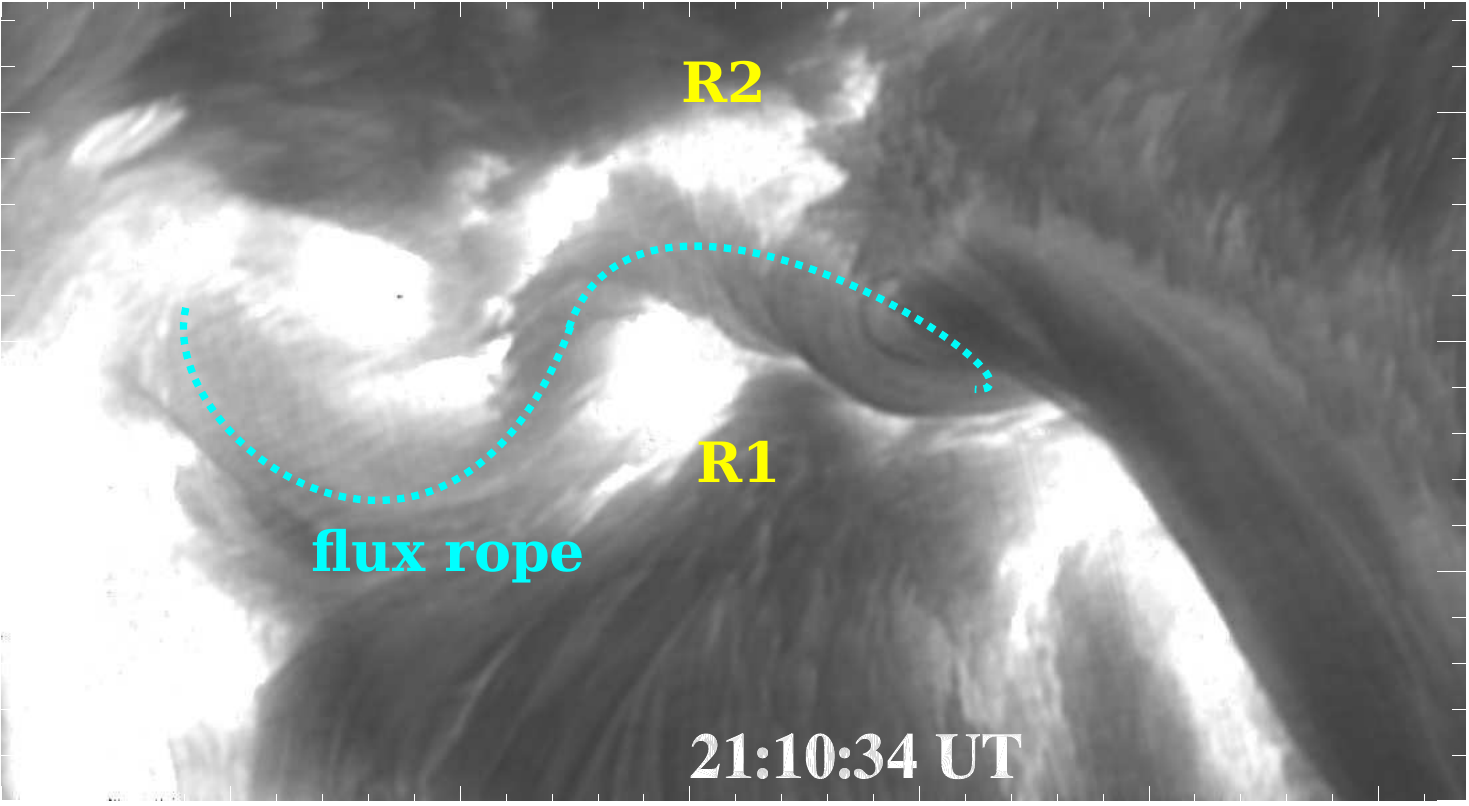}
}
\caption{Selected H$\alpha$ line center images showing a fragment of the rising untwisting jets (yellow ellipse). The size of each image is 32$\arcsec$$\times$17$\arcsec$(each division=1$\arcsec$). L1 and L2 are pre-existing H$\alpha$ loops. Reconnection between these loops probably  generated untwisting jets, a two-ribbon flare (R1 and R2), and appearance of a S-shaped twisted flux rope. Green ellipse in the middle panel indicates the inflow field lines (cool plasma) moving toward the PIL (i.e., reconnection site). (An animation of this figure is available online)}
\label{ha}
\end{figure*}

\begin{figure*}
\centering{
\includegraphics[width=8.1cm]{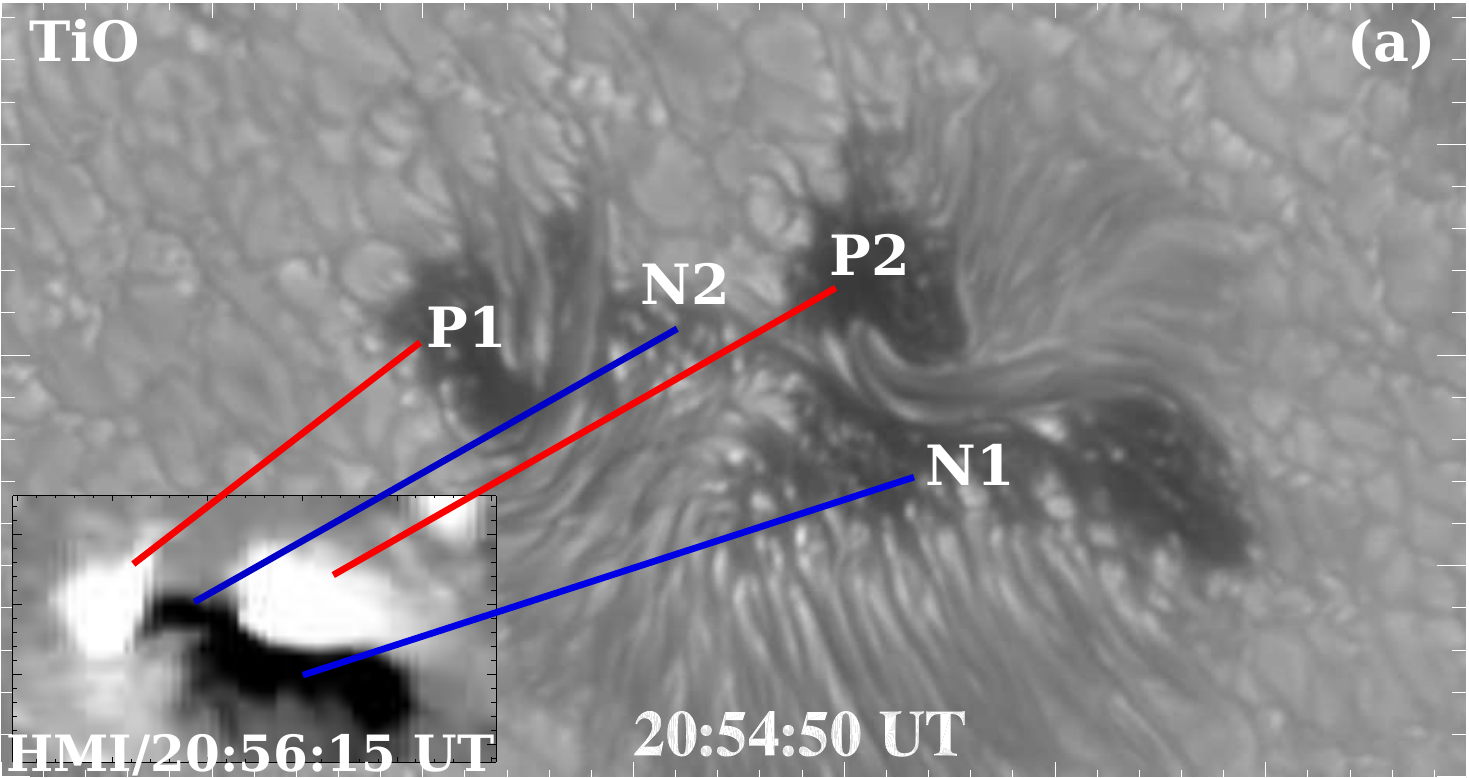}
\includegraphics[width=7.9cm]{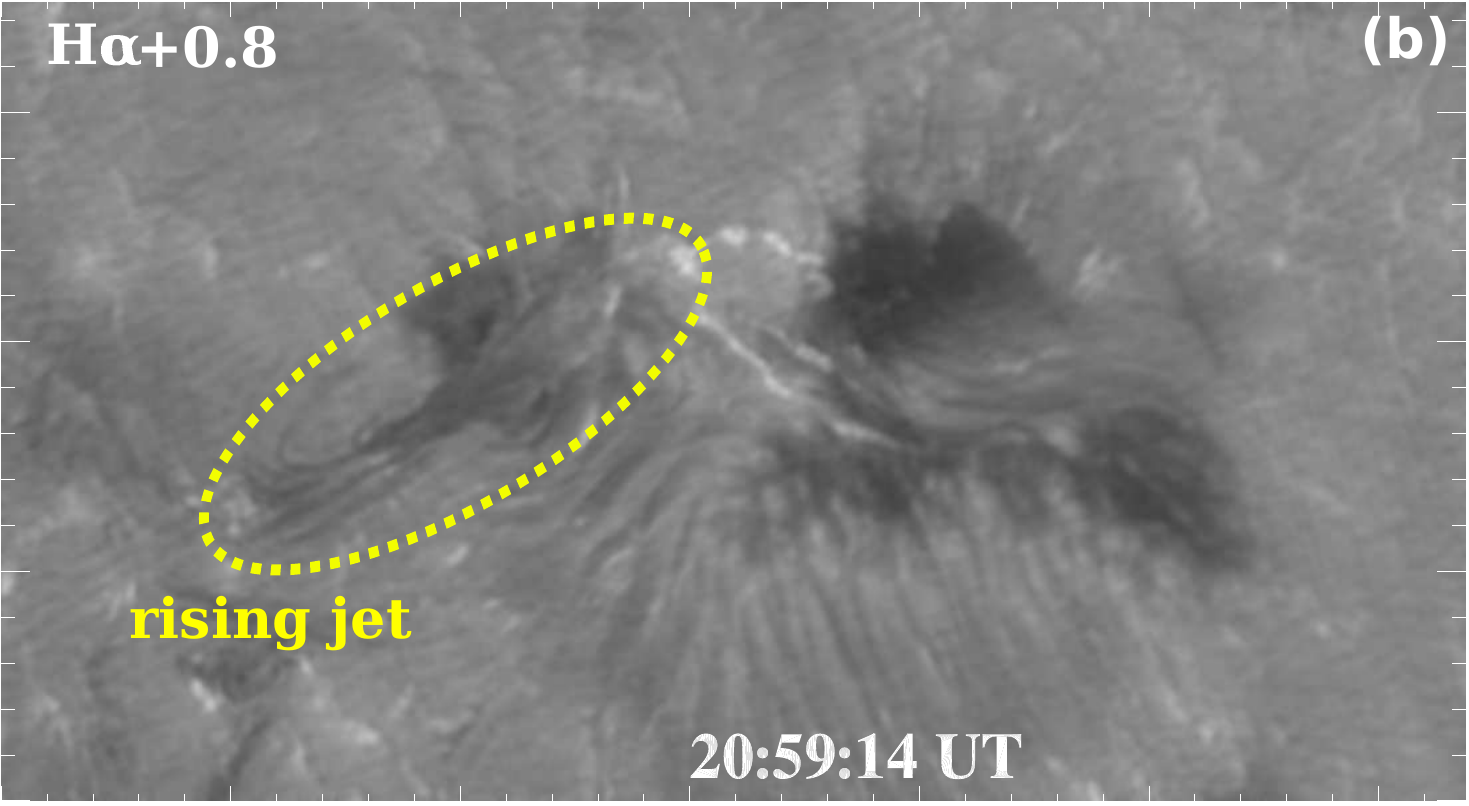}

\includegraphics[width=8cm]{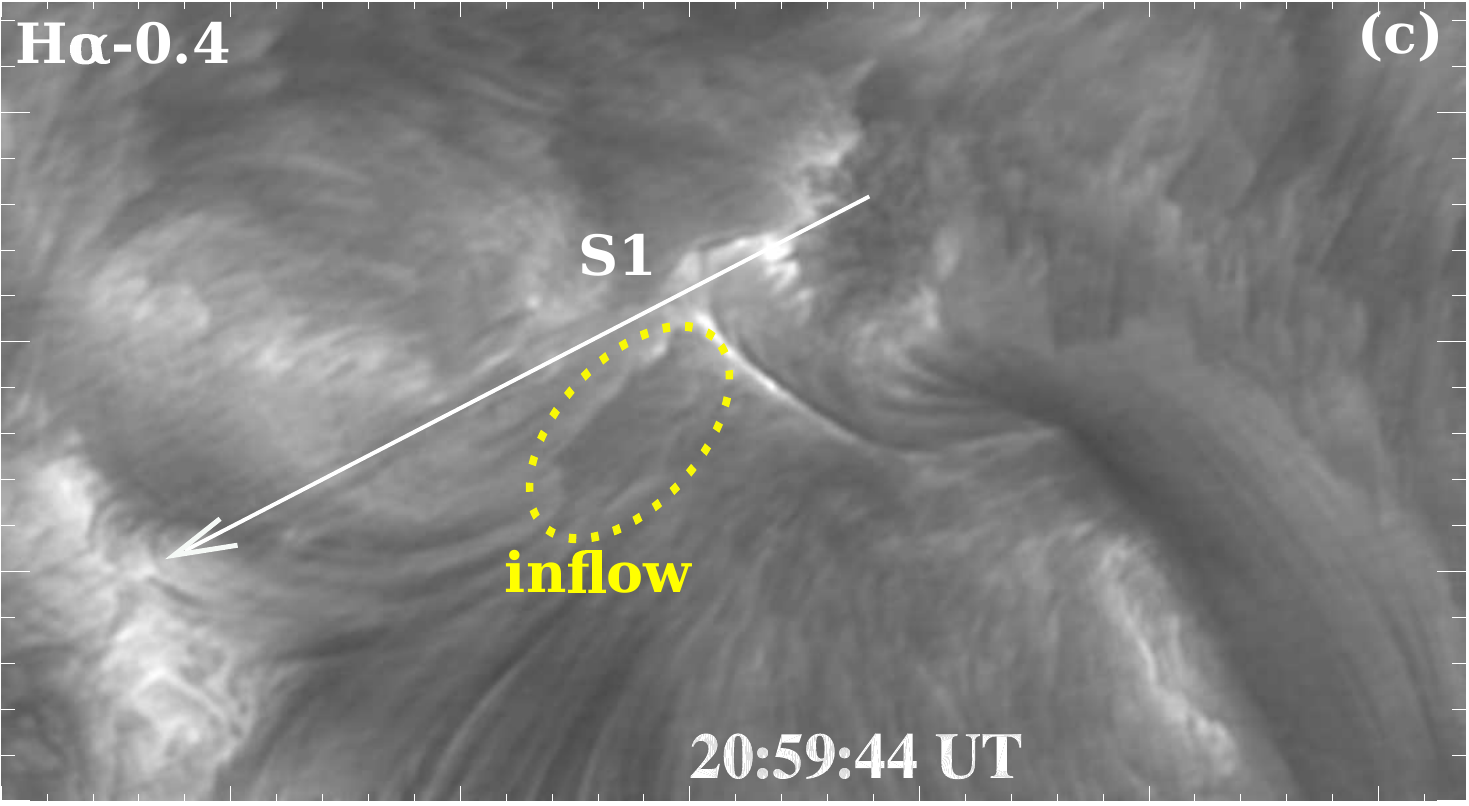}
\includegraphics[width=8cm]{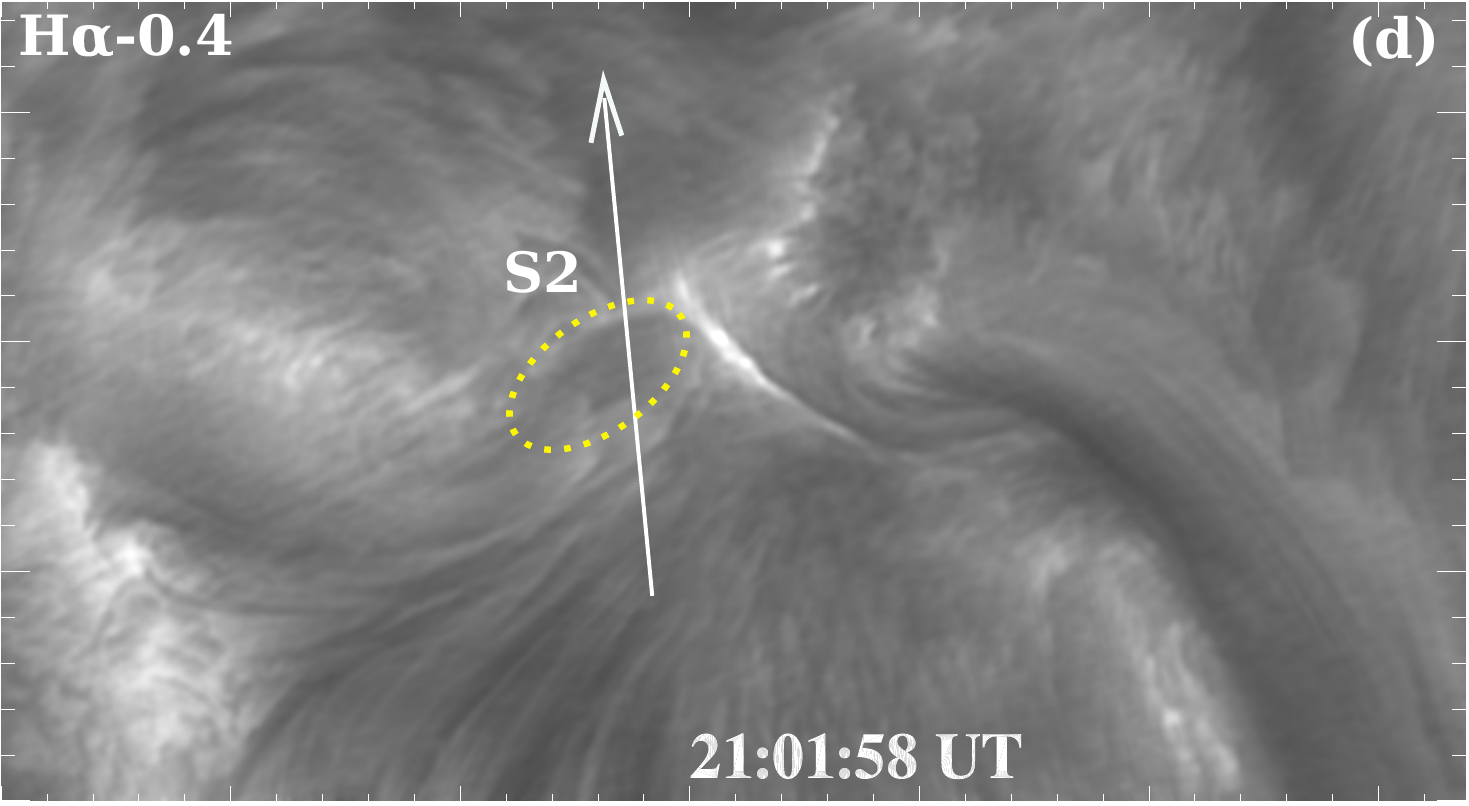}

\includegraphics[width=12cm]{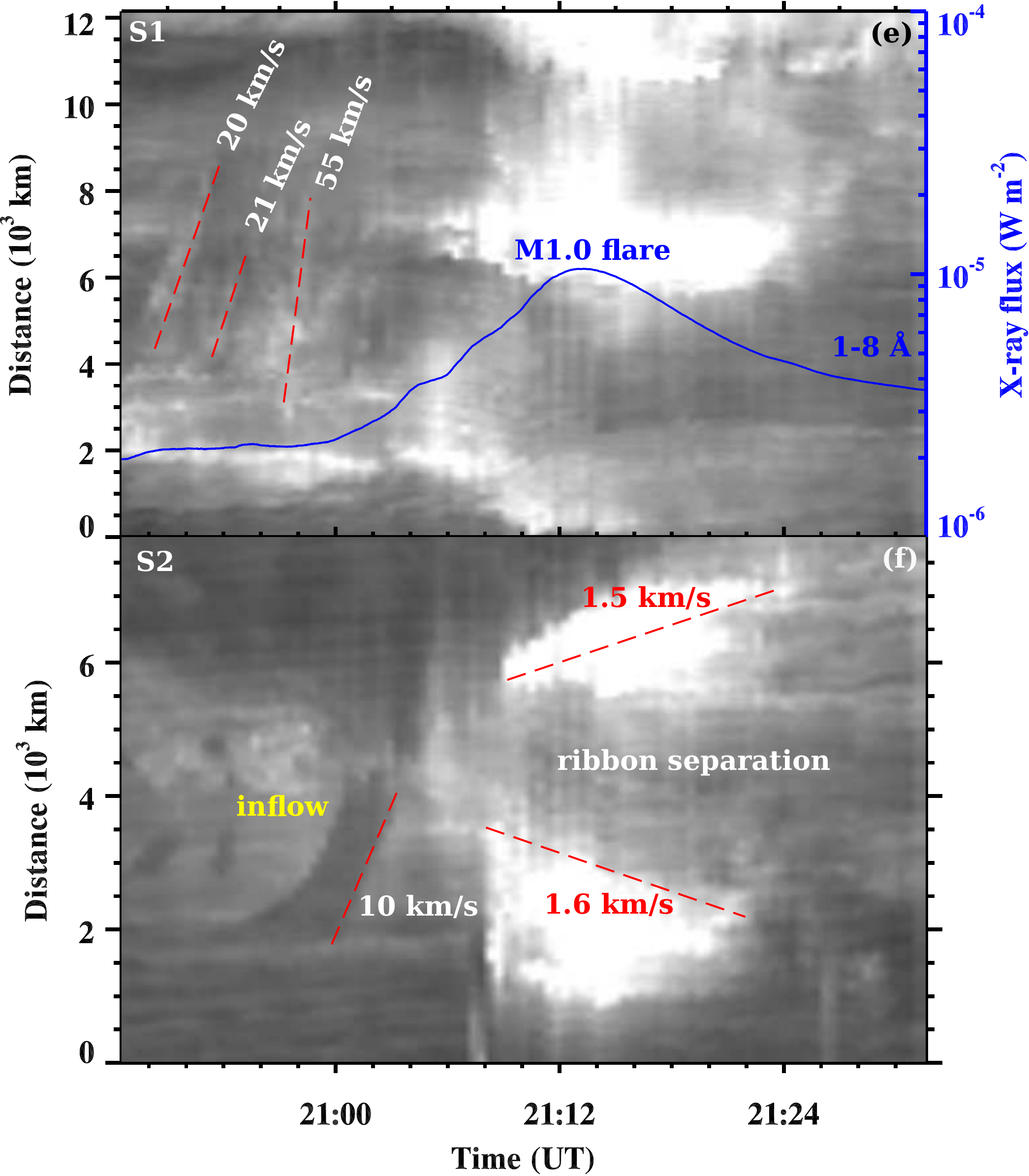}
}
\caption{{\small (a) NST TiO (7057 \AA) image showing the small sunspot group within the AR NOAA 12087. Inset shows the HMI magnetogram image of the same region. P1, P2, N1, and N2 indicate positive and negative polarity sunspots. (b) H$\alpha$+0.8 image showing the rising jet associated with small brightening. (c, d) H$\alpha$-0.4 images of the same region. Yellow ellipse shows the inflow feature at the reconnection site. S1 and S2 are the slice cuts used to create the distance-time plots. 
(e) Distance-time plot of the H$\alpha$-0.4 \AA~ intensity distribution along slice S1 and GOES soft X-ray flux profile (blue) in the 1-8 \AA~ channel. (f) Distance-time plot of the H$\alpha$-0.4 \AA~ intensity distribution along slice S2. (An animation of this figure is available online)}}
\label{hab4}
\end{figure*}


\section{OBSERVATIONS AND RESULTS}
\subsection{Data description}
The NST data were acquired with the aid of the 308 sub-aperture adaptive optics (AO-308).
We used series of narrow-band H$\alpha$ (6563 \AA) images taken at $\pm$0.8 \AA~, $\pm$0.4 \AA~, and 0.0 \AA~ from the line center acquired with NST's Visible Imaging Spectrometer (VIS, pixel size of 0.029$\arcsec$). VIS combines a 5 \AA~ interference filter with a Fabry-P\'erot etalon to produce a resulting bandpass of 0.07 \AA~ over a 70$\arcsec$$\times$70$\arcsec$ field of view. These images provide the view of the different layers of the solar chromosphere. 
Available series of broadband (10 \AA) images of the photosphere were captured with a TiO filter (7057 \AA, pixel scale of 0.0375$\arcsec$). These images are useful to study the evolution of the fine structure of sunspots (for example, flux emergence or cancellation at the photospheric level).

The {\it Atmospheric Image Assembly} (AIA; \citealt{lemen2012}) onboard the {\it Solar Dynamics Observatory} 
(SDO; \citealt{pesnell2012}) 
acquires full disk images of the Sun (field-of-view $\sim$1.3 R$_\odot$) with a spatial resolution of 1.5$\arcsec$ 
(0.6$\arcsec$  pixel$^{-1}$) 
and a cadence of 12 sec in 10 extreme ultraviolet (EUV) and UV channels. This study utilizes 171~\AA\ (Fe IX, $T\approx$0.7 MK), 94~\AA\ (Fe XVIII, $T\approx$6.3 MK), 131 \AA~ (Fe VIII, Fe XXI, Fe XXIII, i.e., 0.4, 10, 16 MK), 304 \AA~(He II, T$\approx$0.05 MK) and 1600~\AA\ (C IV + cont., $T\approx$0.01 MK) images. We also used Heliospheric and Magnetic
Imager (HMI) magnetogram \citep{schou2012} to investigate the magnetic configuration of the active region (AR). 

We also utilized Interface Region Imaging Spectrograph (IRIS, \citealt{depont2014}) slit-jaw images during the flux rope eruption. The pixel size of the slit-jaw images is 0.33$\arcsec$ pixel$^{-1}$ , and a cadence of $\sim$5 s.

\subsection{Chromospheric reconnection and formation of a small flux rope}

The area observed by the NST was lying within the active region (AR) NOAA 12087, which was located at S22E49 on 12 June 2014. The AR was of $\beta\gamma\delta$ magnetic configuration and produced many C \& M-class flares. Two successive flare events (M1.0 and C8.5) are reported here. The first flare was fully observed by the NST, and both flares were covered by IRIS and SDO. The first flare was associated with the untwisting jets,  chromospheric inflow, and a twisted flux rope has appeared (in the NST data) after reconnection. First flare (M1.0) started at $\sim$21:01 UT, peaked at $\sim$21:13 UT, and ended at $\sim$21:19 UT. The second flare (C8.5) began at 00:30 UT, maximized at 00:34, and ended at 0:41 UT. Note that an M3.1 flare was also observed between our studied two flares, which originated from a different active region (AR 12085)

Figure \ref{ha} displays some of the selected H$\alpha$ images recorded by the NST. Before the onset of the first M1.0 flare, we observe two H$\alpha$ cool loops (J shaped) at 20:48 UT, which are indicated by L1 (red) and L2 (blue). These loops were rooted in opposite polarity sunspots and lying along the PIL. Later at $\sim$20:50 UT,  we already can see small brightening between these loops, followed by the rise of small jet-like structures with untwisting motions. An H$\alpha$ movie shows multiple jet structures that emanate from the brightening site between L1 and L2. The rise of the untwisting jets drives cool plasma inflow (green ellipse) behind it at $\sim$20:59 UT. The inflow of field lines was possibly caused by the evacuation due to reduced plasma density (i.e., pressure) behind the erupting small untwisting jets. This is an example of a unidirectional cool plasma inflow. The distance time plot clearly shows that the inflow direction is toward the PIL, and the direction of the ribbon separation is across the PIL, which is in agreement with the previous observations \citep{takasao2012,kumar2013inflow,su2013}. In addition, the inflow is co-spatial with the location of the joining of two chromospheric loops  L1 and L2. When these oppositly directed field lines reconnect, impulsive energy release starts. The chromospheric inflows initiated magnetic reconnection, and a two-ribbon flare (M1.0) progressed from $\sim$21:04 UT onward. R1 and R2 indicate the flare ribbons. Interestingly, we observed appearance of a twisted flux rope (S-shaped) along the PIL as a result of magnetic reconnection (marked by dashed line at 21:10 UT). The coalescence of loops L1 and L2 was clearly seen in the H$\alpha$ movies during magnetic reconnection. Therefore, the flux rope has most likely formed during the magnetic reconnection between two cool loops L1 and L2. 

Figure \ref{hab4}(a-d) display selected images of the flare taken in TiO (7057 \AA), H$\alpha$+0.8, and H$\alpha$-0.4 bands as observed by the NST. Figure \ref{hab4}(a) shows the photospheric image exhibiting the fine structure of the small sunspots before the flare onset at 20:54:50 UT. To compare the polarities of these spots, we display a HMI line-of-sight magnetogram in the lower left corner. P1, P2, N1, and N2 represent positive and negative polarity sunspots. N1 and P2 comprise a delta type magnetic configuration. A part of the rising jet-like structure is shown in Figure \ref{hab4}(b), which is associated with small chromospheric brightening at the footpoint between P1 and N1. Figure \ref{hab4}(c,d) display cool chrmospheric plasma inflows (dashed ellipse) at the reconnection site between opposite polarity sunspots (i.e., toward the PIL).

To investigate the signature of magnetic reconnection before the onset of the M1.0 flare, we created a stack plots of these H$\alpha$-0.4 \AA~ intensity slices made along the cuts S1 and S2. Figure \ref{hab4}(e,f) displays space-time plots. The GOES soft X-ray flux profile is overplotted in panel (e) to compare the time of untwisting jets and associated plasma inflow. The rise of small jets was observed during 20:48-20:56 UT. The speed of the jets (from the linear fit) ranges from $\sim$20-55 km s$^{-1}$. The rise of the jets induced plasma inflows at the reconnection site during $\sim$20:54-21:02 UT (Figure \ref{hab4}(d)). The inflow speed was $\sim$10 km s$^{-1}$. After plasma inflow, we notice the flare brightening in the chromosphere during its impulsive phase. It is similar to a classical two-ribbon flare. The ribbon separation speed is $\sim$1.5-1.6 km s$^{-1}$. Note that in strong eruptive flares, the ribbon separation speed can be $\sim$10-15 km s$^{-1}$ \citep{wang2003}.

\begin{figure*}
\centering{
\includegraphics[width=8cm]{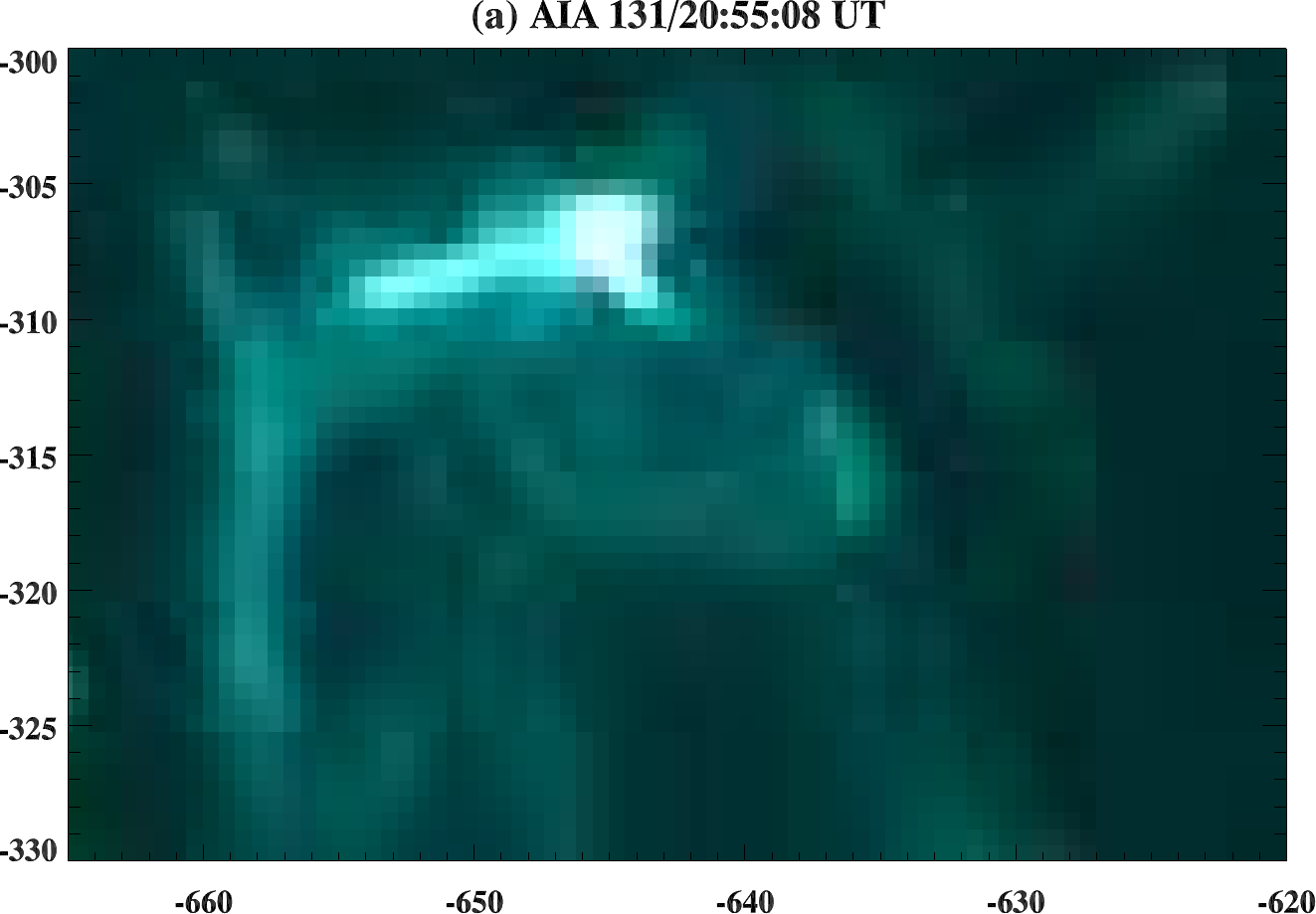}
\includegraphics[width=8cm]{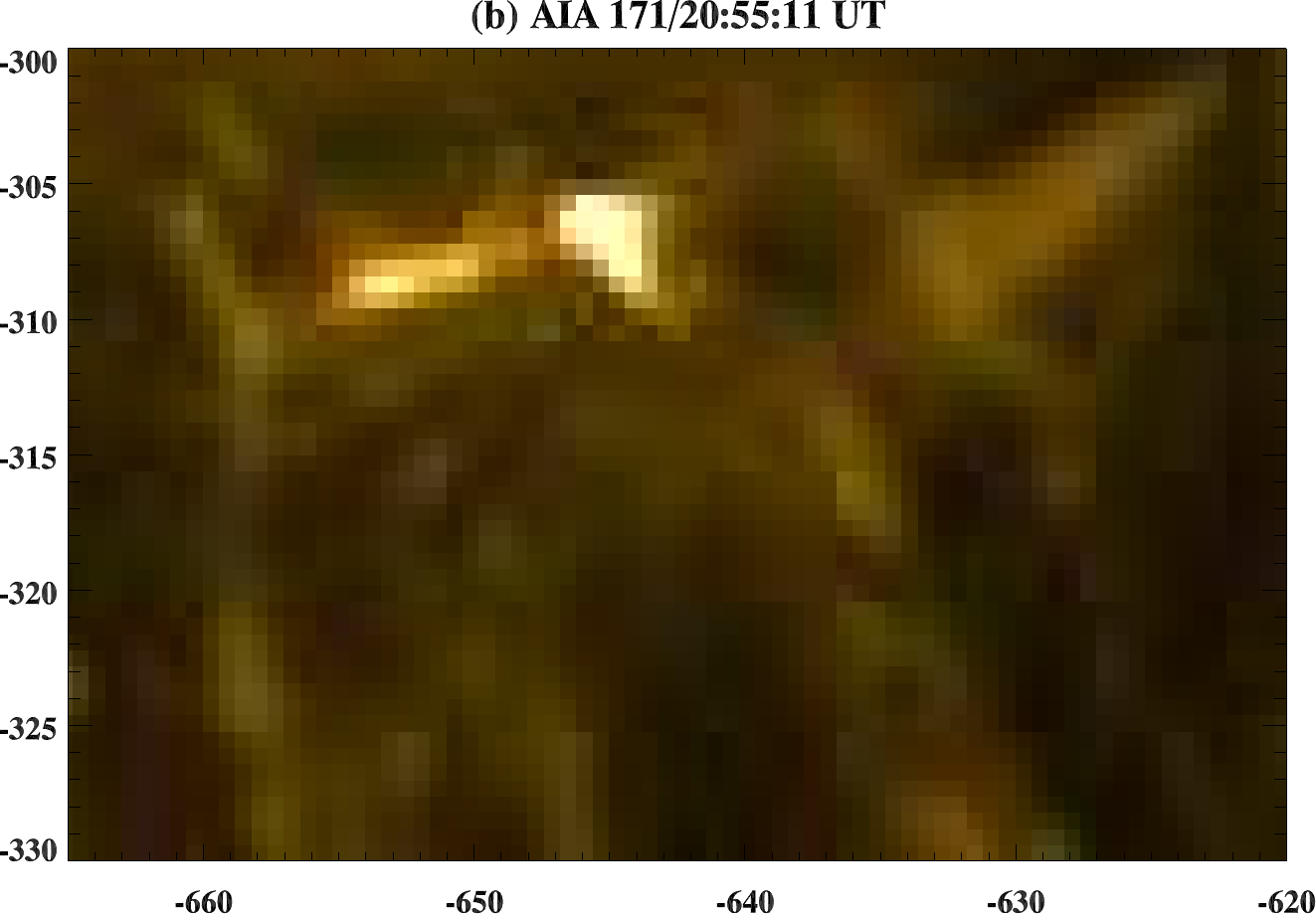}

\includegraphics[width=8cm]{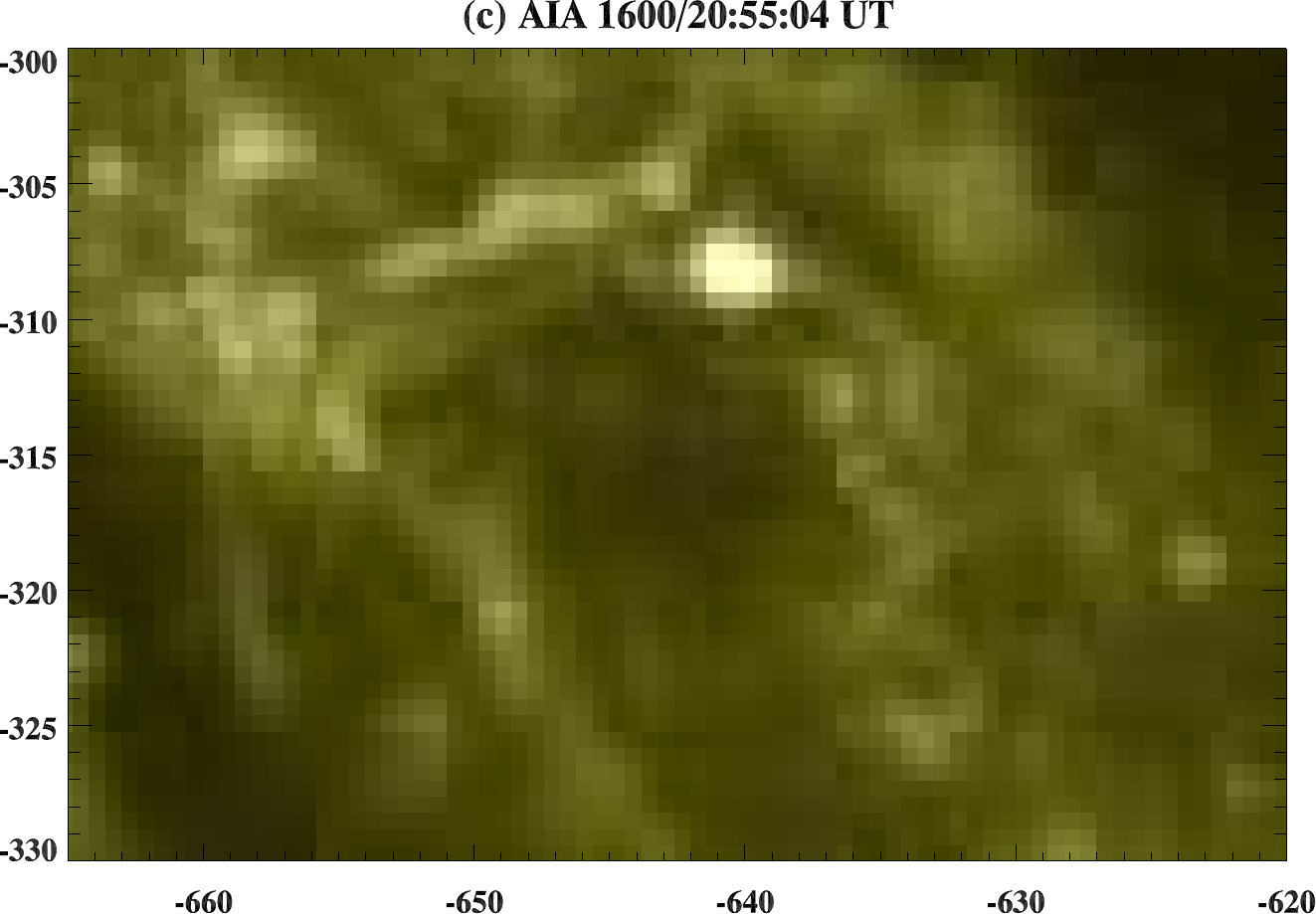}
\includegraphics[width=8cm]{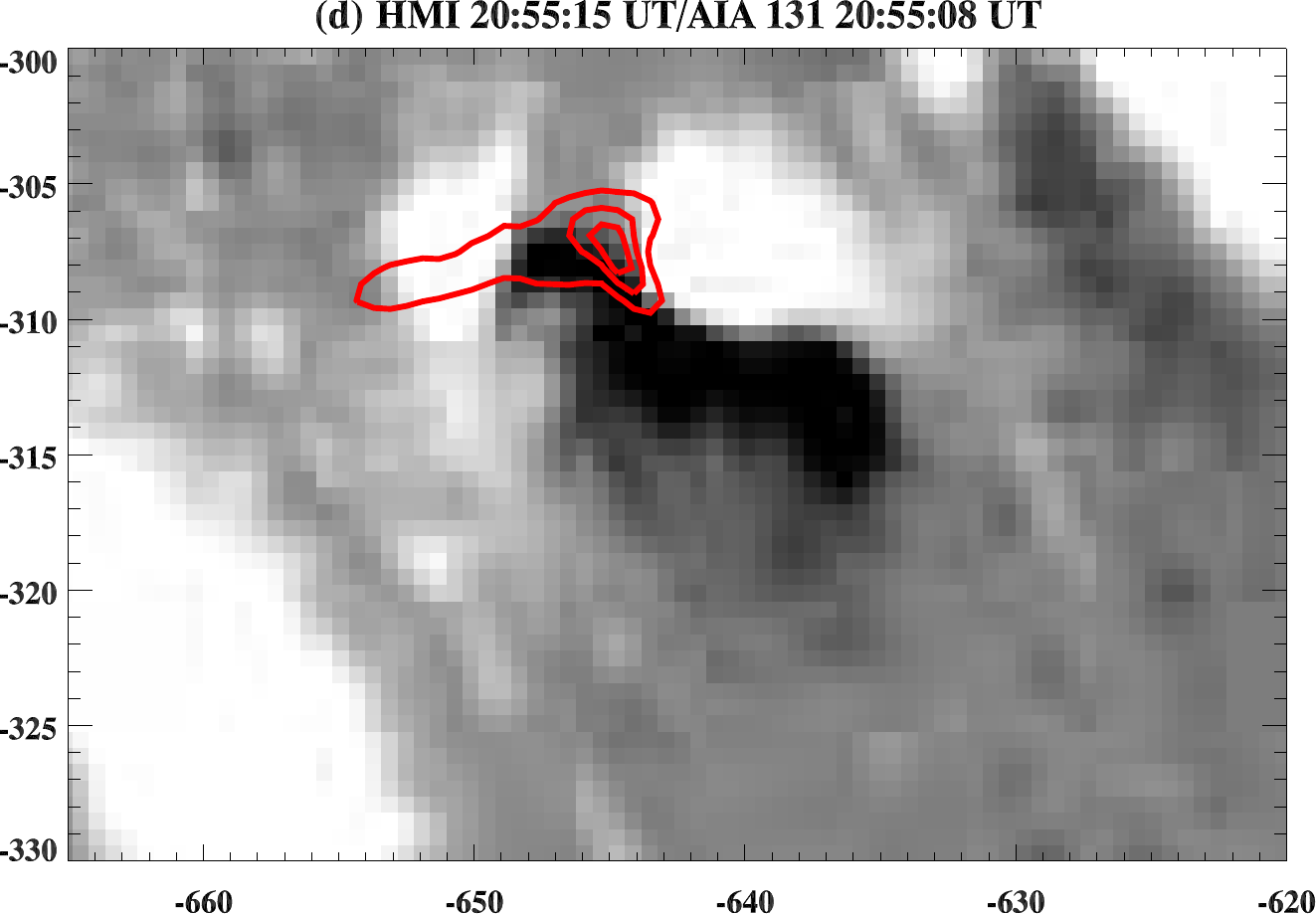}

}
\caption{(a,b,c) AIA 131, 171, and 1600 \AA~ images during the preflare phase (20:55 UT). (d) HMI magnetogram overlaid by EUV brightening observed in the AIA 131 \AA~. X and Y axis of each image are in arcsecs.}
\label{preflare}
\end{figure*}

\begin{figure*}
\centering{
\includegraphics[width=6cm]{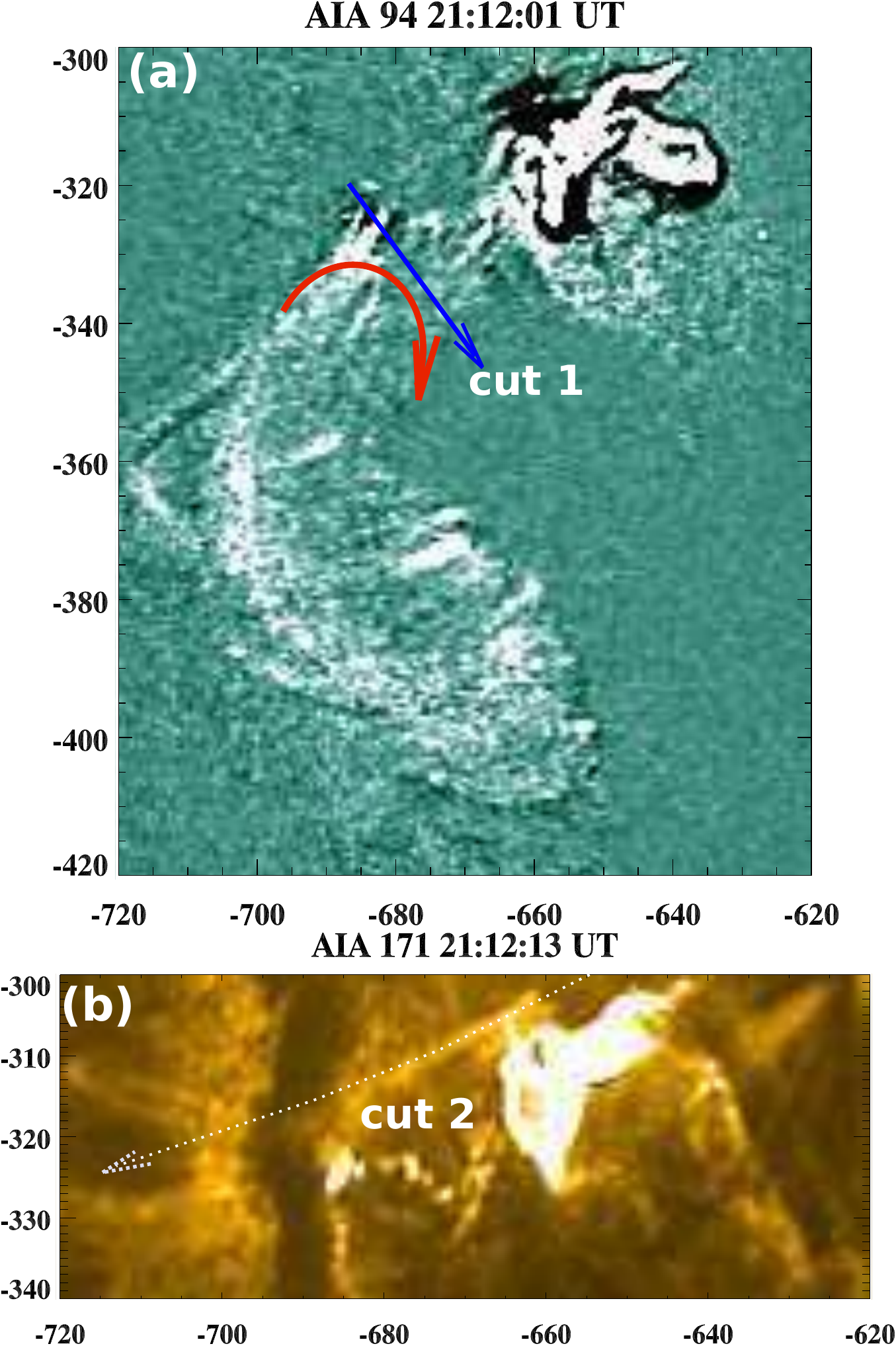}
\includegraphics[width=8.5cm]{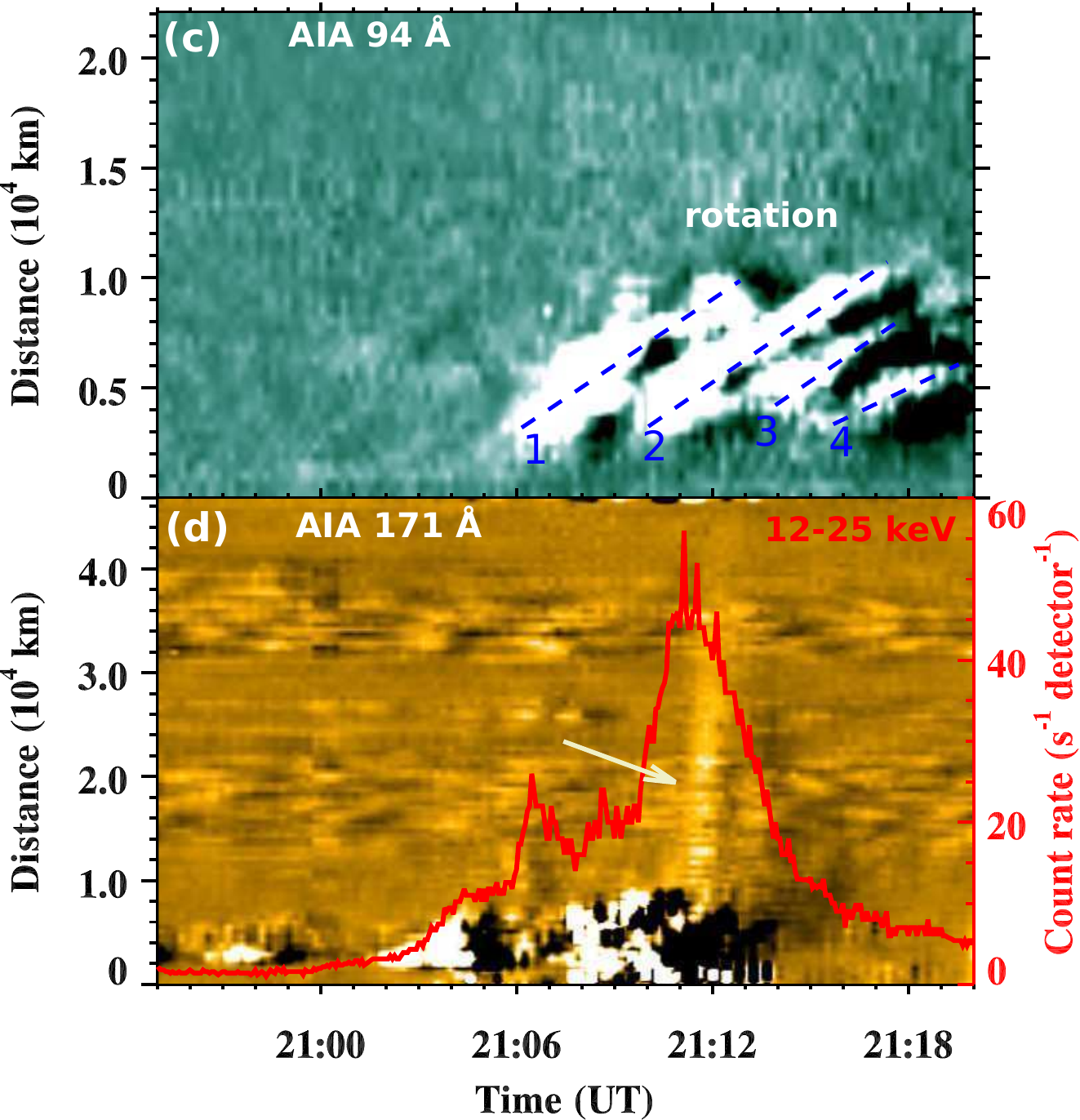}
}
\caption{(a-b) AIA 94 and 171 \AA~ images during the flare maximum phase ($\sim$21:12 UT). The red arrow shows the direction of rotation of the brightening patches. (c-d) Distance-time plots of intensity distribution along cut 1 and cut 2 using AIA 94 and 171 \AA~ running difference images. The red curve shows the RHESSI hard X-ray flux profile in 12-25 keV energy channel. The plasma flow during the slippage of the 171 \AA~ loops is marked by an arrow. (An animation of this figure is available online)}
\label{rotation}
\end{figure*}


\subsection{Preflare brightening and coronal response}

As we mentioned before, during the preflare phase we observed rise of untwisting jets rise in the H$\alpha$ images. To examine the coronal activity associated with the chromospheric reconnection, we display AIA images in 131, 171 and 1600 \AA~ at $\sim$20:55 UT (Figure \ref{preflare}(a,b,c)). We clearly notice plasma heating at the footpoint of rising jet, where the chromospheric reconnection occurred between two sheared H$\alpha$ loops. These images  also show the elongated  jet, however not so clear due to low spatial resolution. We overlaid AIA 131 \AA~ brightening over a co-temporal HMI magnetogram (Figure \ref{preflare}(d)). The plasma heating occurred exactly at the site of joining of two sheared loops in the chromosphere. 

In Figure \ref{rotation} (a,b), we display AIA 94 running difference and 171 \AA~ intensity images. Cut 1 and cut 2 are the slices used to create the stack plots.
 Interestingly, AIA 94 \AA~ movie reveals the apparent rotation of the brightening patches (clockwise direction, marked by red arrow) above the flare site during 21:06-21:20 UT (Figure \ref{rotation}(c)). Hot loops observed in the AIA 94 \AA~ channel were connected to the flare site. The rotation of the brightening patches was followed by the apparent slippage of the (southward) 171 \AA~ loops (refer to AIA 94 and 171 \AA~ movie). A cool plasma flow (171 \AA) was also observed along the field lines during the slippage motion (Figure \ref{rotation}(d)). The apparent linear speed (v) of the brightening patches in the sky plane (marked by 1, 2, 3, 4) is 16.6, 17, 16.8, and 11.8 km s$^{-1}$, respectively. Using the approximate width of the rotation region as a diameter (from the 94 \AA~ stack plot) $\sim$7$\times$10$^3$ km, we can calculate the radius (r)  $\sim$3.5$\times$10$^3$ km. If we use the average linear speed of the field lines $\sim$15.5 km s$^{-1}$, the estimated apparent rotational speed ($\omega$=v/r) will be $\sim$15 degree min$^{-1}$. This is the lower limit of the projected rotational speed. The rotation of the brightening patches around the spine implies the apparent slippage of magnetic field lines and may be an additional evidence of fan-spine topology. According to the theory of the 3D torsional spine reconnection \citep{pontin2007,priest2009}, the rotational slippage of field lines around the spine should be observed when the fan drives torsional spine reconnection with a strong spine current. Alternatively, apparent sub-Alfv\'enic motion of the field lines (or sequential brightening) may also be interpreted as a result of slipping magnetic reconnection around null-point as predicted in the numerical simulations \citep{aul2006,masson2009,torok2009}. 
The rotation of the sequential brightening patches shows good correlation with the hard X-ray burst (12-25 keV, red curve), suggesting the ongoing reconnection and associated particle acceleration.

\begin{figure*}
\centering{
\includegraphics[width=16cm]{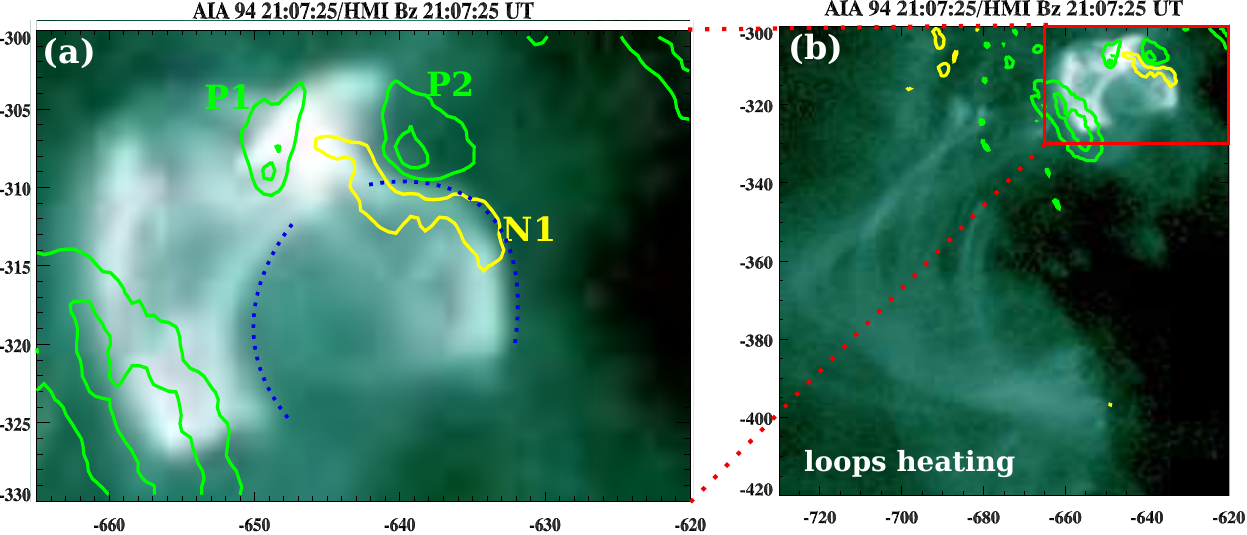}

\includegraphics[width=8cm]{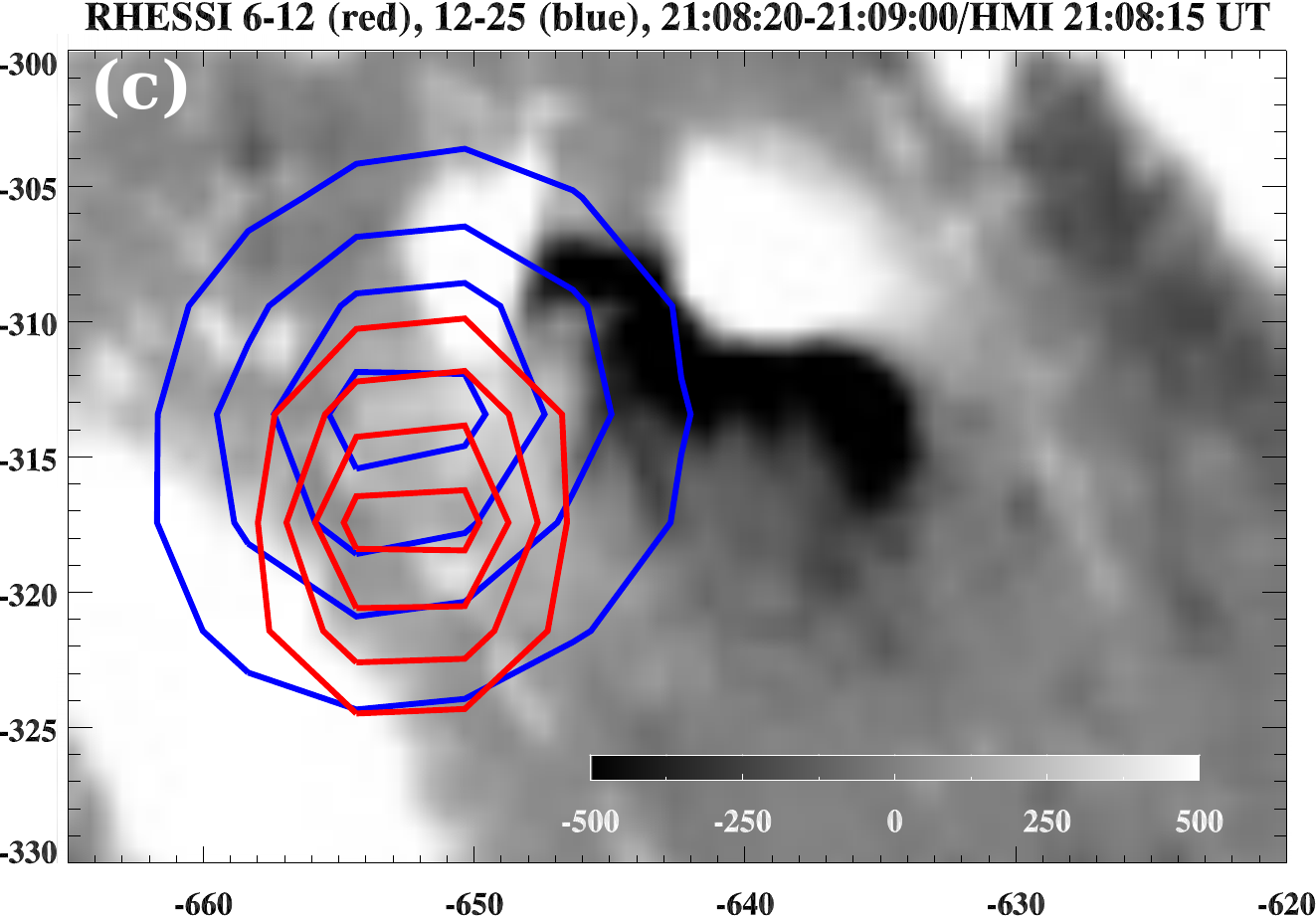}
\includegraphics[width=8cm]{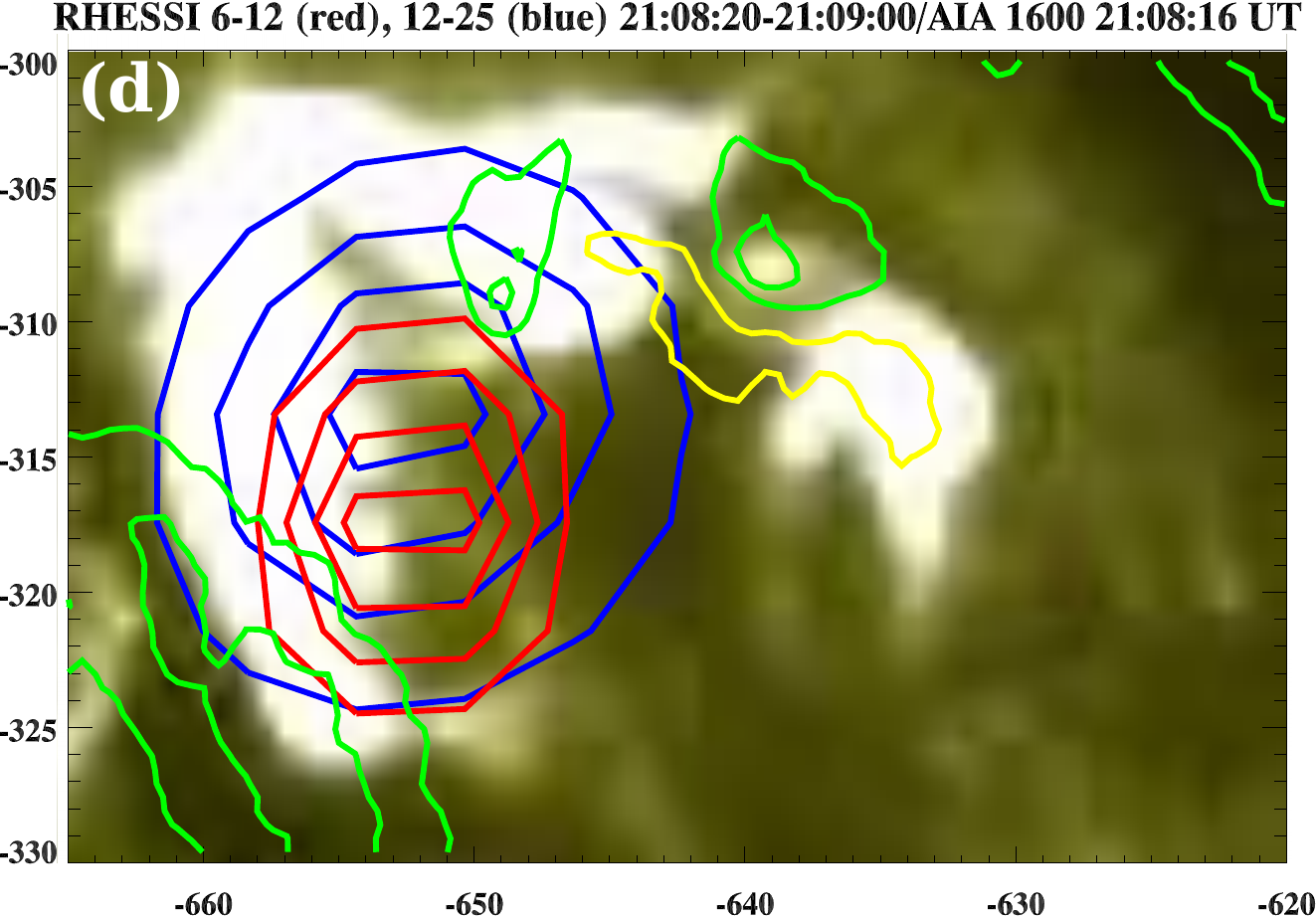}

\includegraphics[width=8cm]{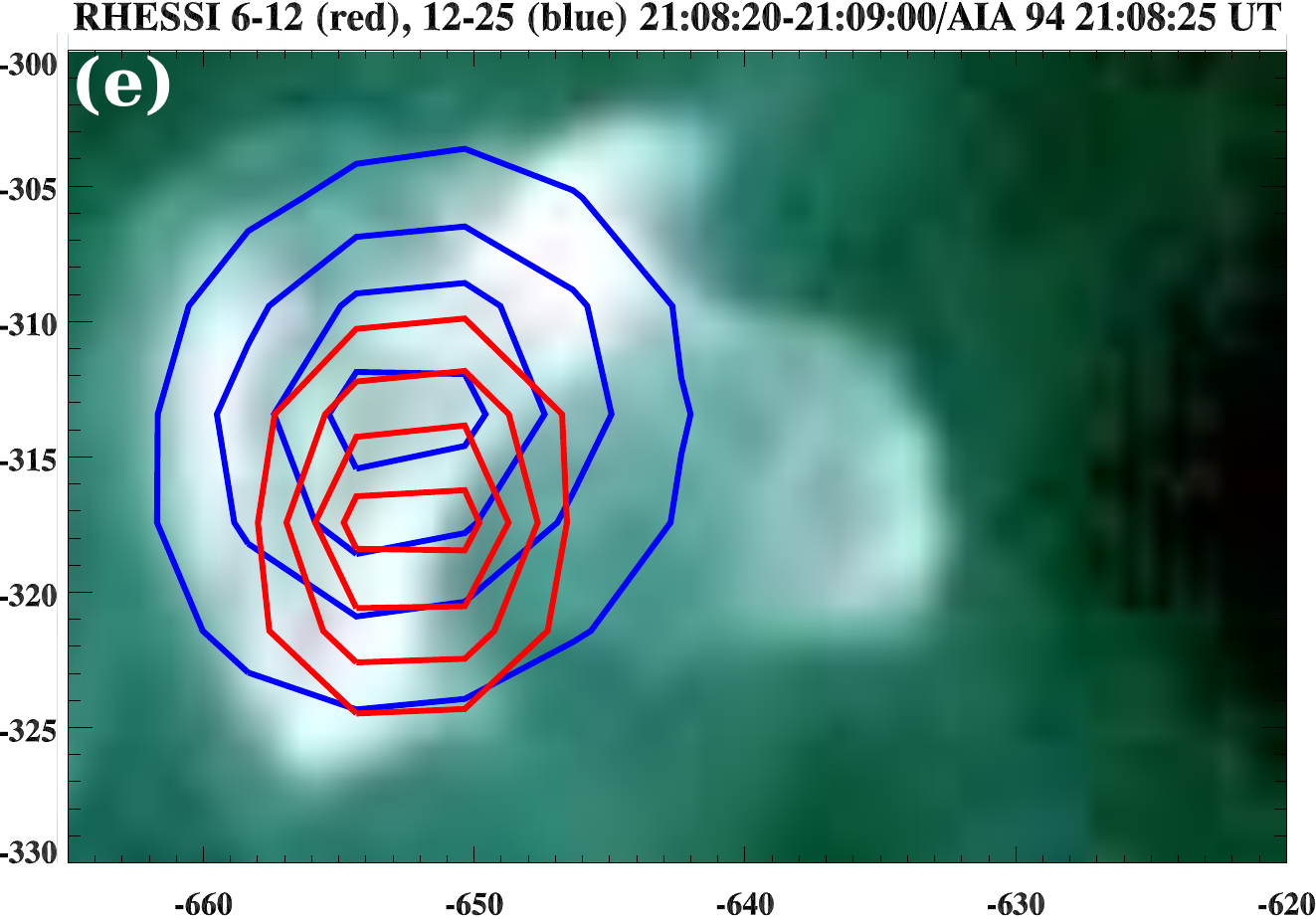}
\includegraphics[width=8cm]{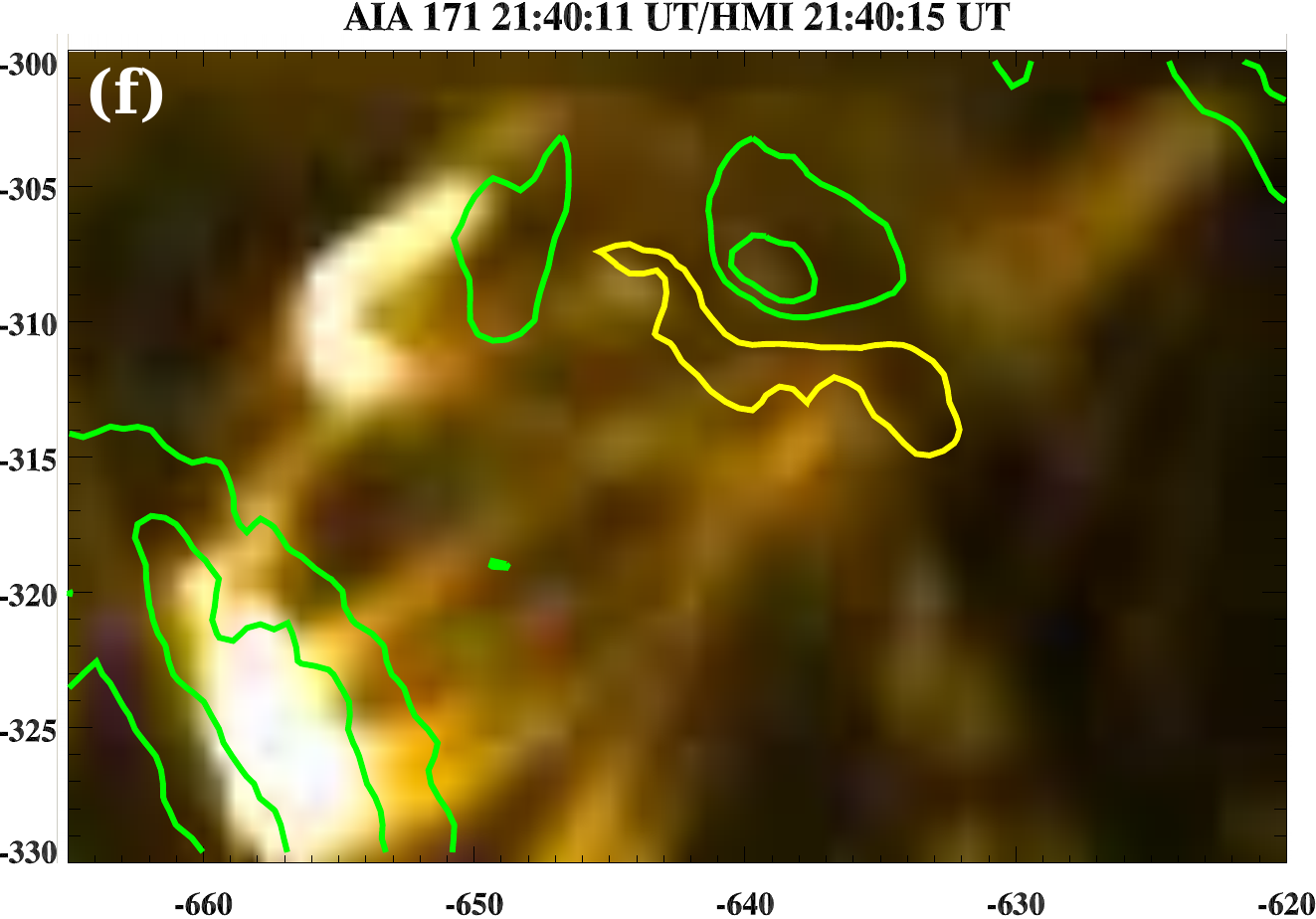}

}
\caption{(a,b) AIA 94 \AA~ images overlaid by HMI magnetogram contours of positive (green) and negative (yellow) polarities. Red box shows the field of view of panel (a). Blue dotted lines indicate the connectivity of the field lines. (c,d,e,f) RHESSI X-ray contours (6-12 keV in red, 12-25 keV in blue) overlaid on HMI magnetogram, AIA 1600 and 94 \AA~ images during flare impulsive phase. Contour levels are 30$\%$, 50$\%$, 70$\%$, and 90$\%$ of the peak intensity. HMI magnetogram contours of positive (green) and negative (yellow) polarities are overlaid on AIA 1600 (during flare) and 171 (after flare) \AA~ images. Contour levels are $\pm$500, $\pm$1000, and $\pm$1500 Gauss. X and Y axis of each image are in arcsecs. (An animation of this figure is available online)}
\label{hessi}
\end{figure*}

To examine the structure of coronal magnetic fields and coronal response to the flare, we used AIA images in 94, 171, and 1600 \AA~ channels. Figure \ref{hessi}(a,b) displays AIA 94 \AA~ images during the flare impulsive phase (21:07 UT). These images are overlaid by HMI magnetogram contours of positive (green) and negative (yellow) polarities. The coronal view of the field structure is revealed once the heated plasma from the footpoints is pumped and filled these field lines (Figure \ref{hessi}(a)). The enlarged field of view shows the heating of a remote loop during the flare impulsive phase. One footpoint of the heated loop is rooted near the main flare site. The acceleration of energetic particles from the flare site possibly heats this loop system. The accelerated particles can be confined to the field line, and precipitate at the opposite footpoint of the large loop observed in the hot AIA channels. 

We analyzed RHESSI hard X-ray images \citep{lin2002} to explore the evolution of the hard X-ray sources during the flare. We adopted the PIXON algorithm technique \citep{metcalf1996} for the image reconstruction. We used 40 s integration time in both energy channels. 
A single hard X-ray source (6-12, 12-25 keV) was observed 21:05 UT onward. Figure \ref{hessi}(c,d,e) displays the RHESSI hard X-ray contours overlaid on the HMI magnetogram, AIA 1600 and AIA 94 \AA~ images during the flare impulsive phase (21:08 UT). The formation of a single hard X-ray source reveal a loop-top source (particle acceleration site) in the corona. The post-flare connectivity of the magnetic field lines is shown by AIA 171 image at 21:40 UT (Figure \ref{hessi}(f)). This image shows the connectivity of negative polarity spot (N1) with the surrounding opposite polarity fields. Note that global ribbon morphology is quasi-circular in the AIA 1600 \AA~ image. Flare onset begins at the site of the emerging small untwisting jets, and most likely these jets not only induce reconnection behind in the chromosphere but also destabilize the overlying fields in the corona. An extended quasi-circular ribbon formed as a result of particle precipitation from the acceleration site in the corona. The circular ribbons are usually formed due to fan-spine topology of the magnetic field configuration \citep{masson2009}. 

 The global morphology of the flare ribbon is quasi-circular. The NST high resolution observations clearly reveal the ribbon separation (locally) associated with the rising motion of the untwisting jet in the initial phase of the flare. This mechanism most likely resembles the ribbon separation as a result of magnetic reconnection occurring beneath an erupting filament/flux rope. 
Very likely, the reconnection of the untwisting jet with the overlying field (possibly at the coronal null-point) resulted in particle acceleration along the fan loops to form a quasi-circular ribbon. However, the ribbon separation (within global quasi-circular ribbon) associated with the eruption of an untwisting jet might not have been observed before because of low spatial resolution of the previous data sets. In addition, the ribbon separation speed is not much as high as is usually observed in the filament eruptions.

The reason why we consider a fan-spine topology (without magnetic field extrapolation) is due to the following observational evidences:
(i) formation of the global quasi-circular flare ribbon during both flares \citep{masson2009}, (ii) apparent rotation/slippage of the field lines (i.e., sequential brightening) around the outer spine during the first flare \citep{pontin2007,priest2009}.  Therefore, we speculate the existence of a coronal null point associated with the fan-spine topology.

\subsection{Flux rope eruption}
\begin{figure*}
\centering{
\includegraphics[width=5.3cm]{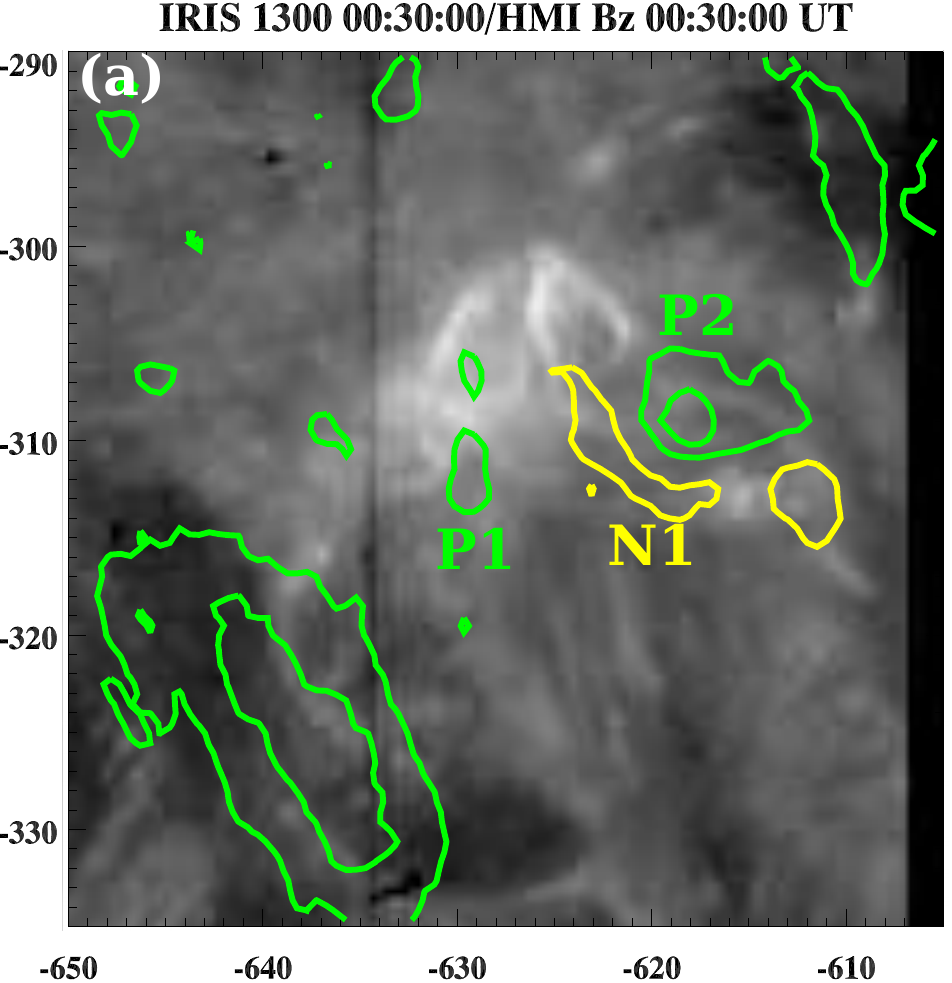}
\includegraphics[width=5.3cm]{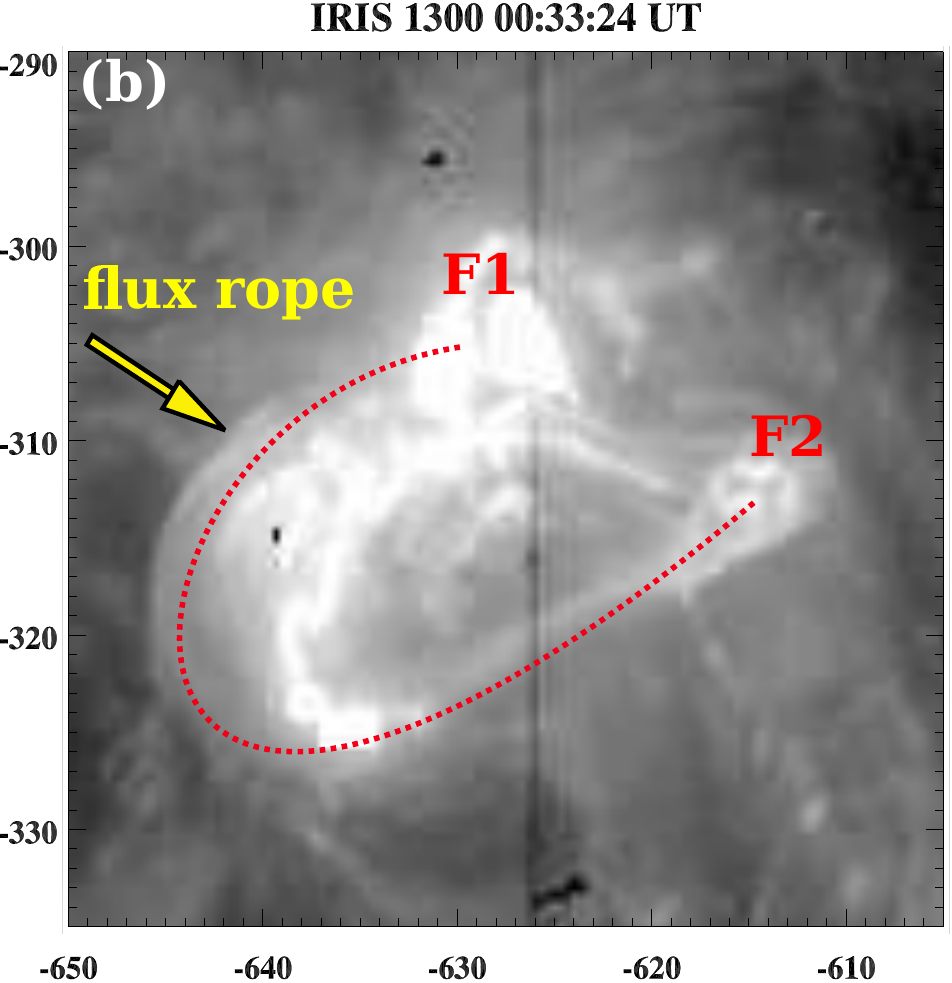}
\includegraphics[width=5.3cm]{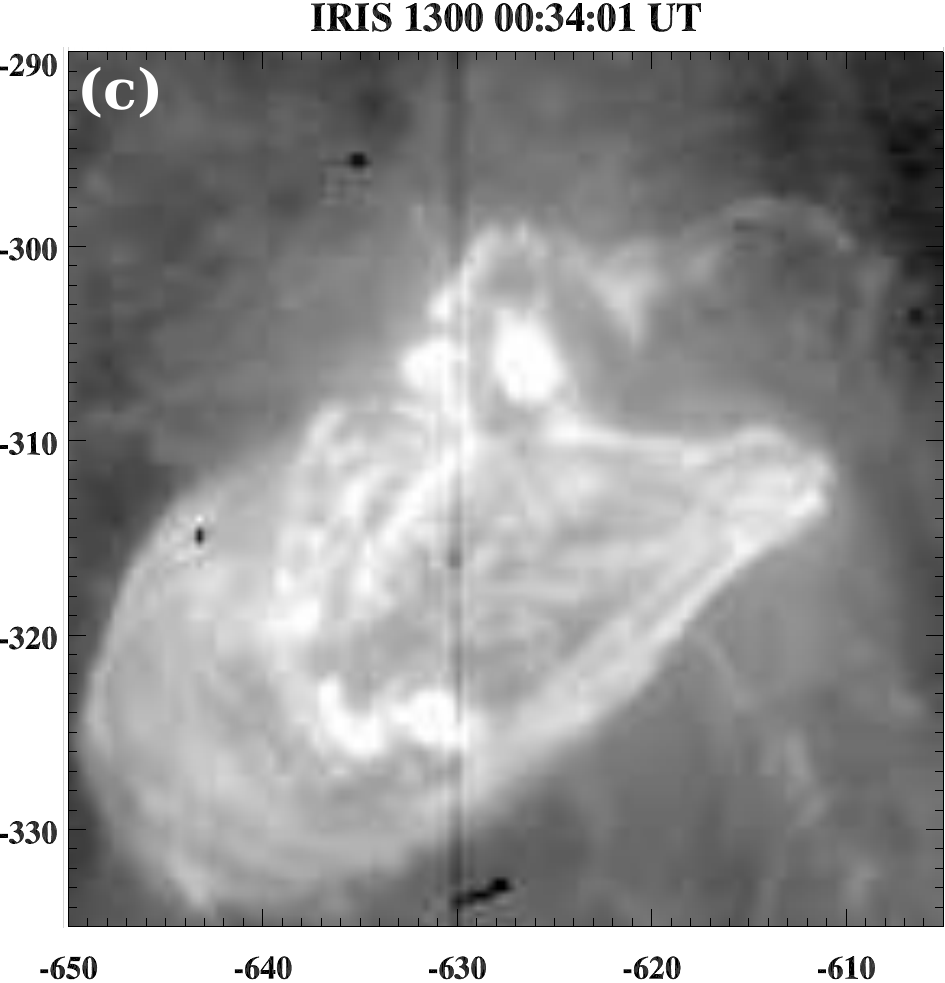}

\includegraphics[width=5.3cm]{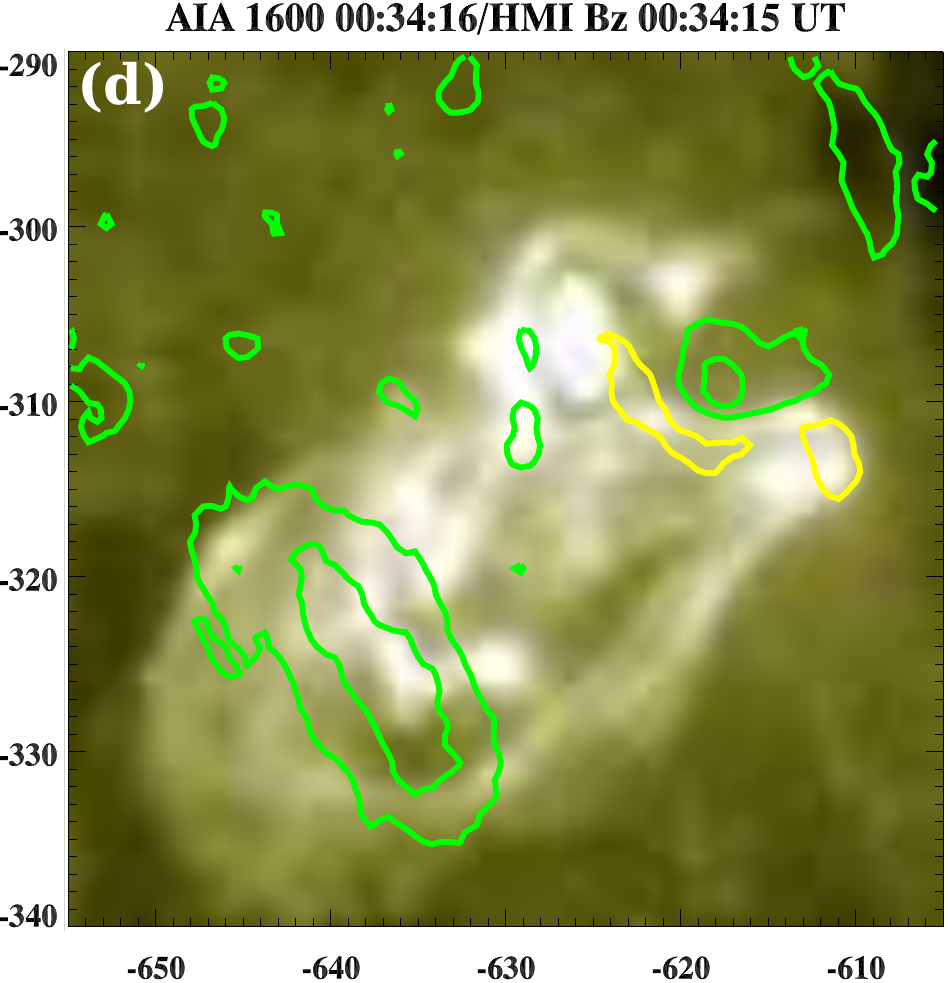}
\includegraphics[width=5.3cm]{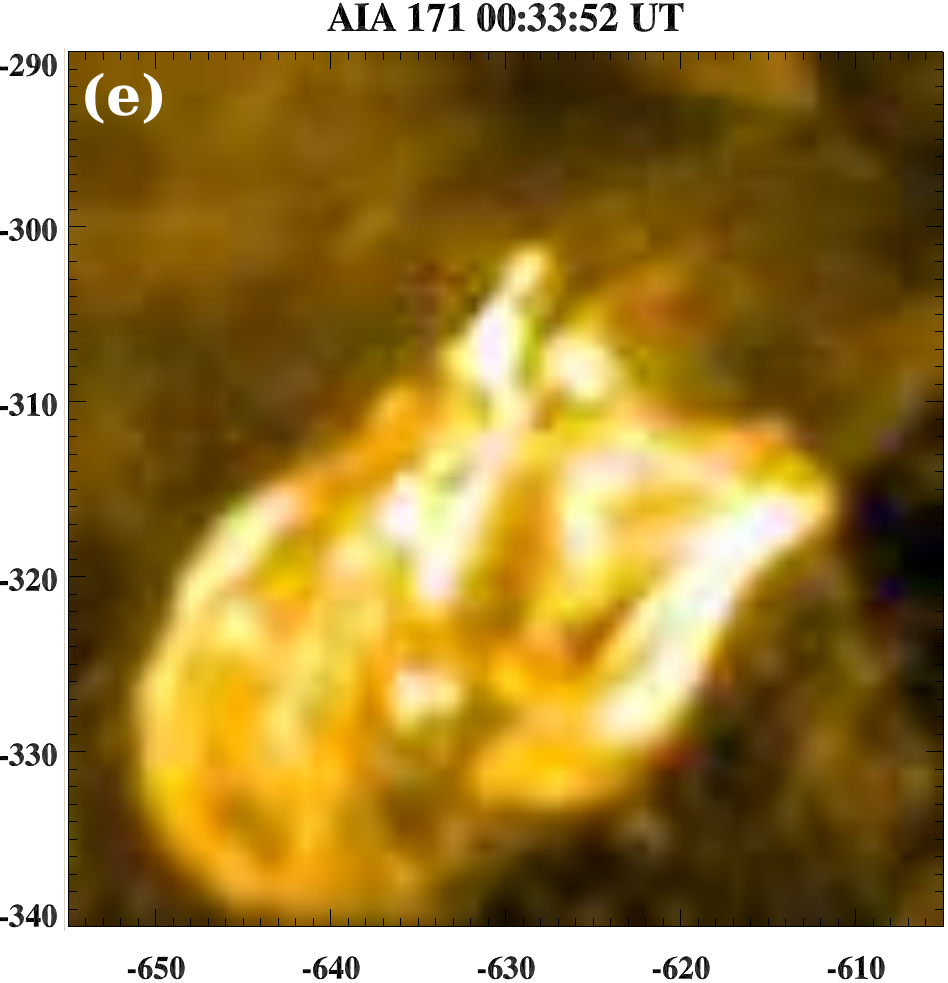}
\includegraphics[width=5.3cm]{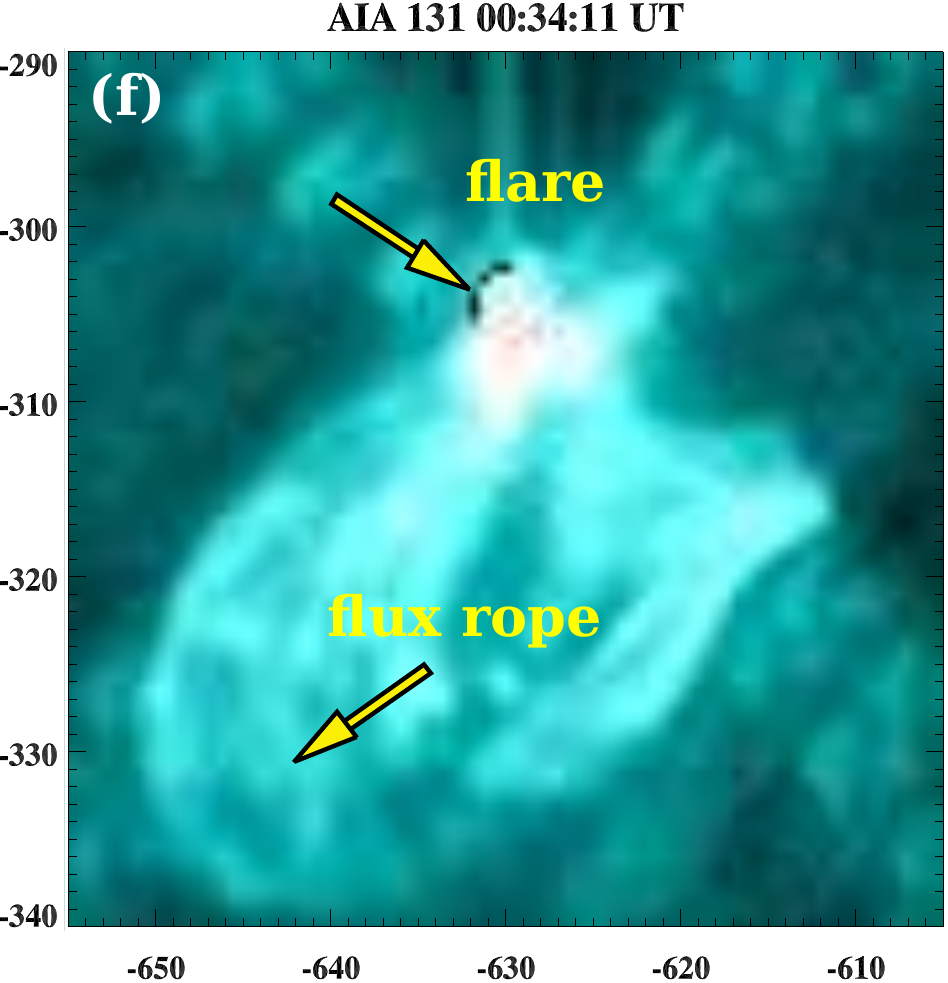}

\includegraphics[width=5.3cm]{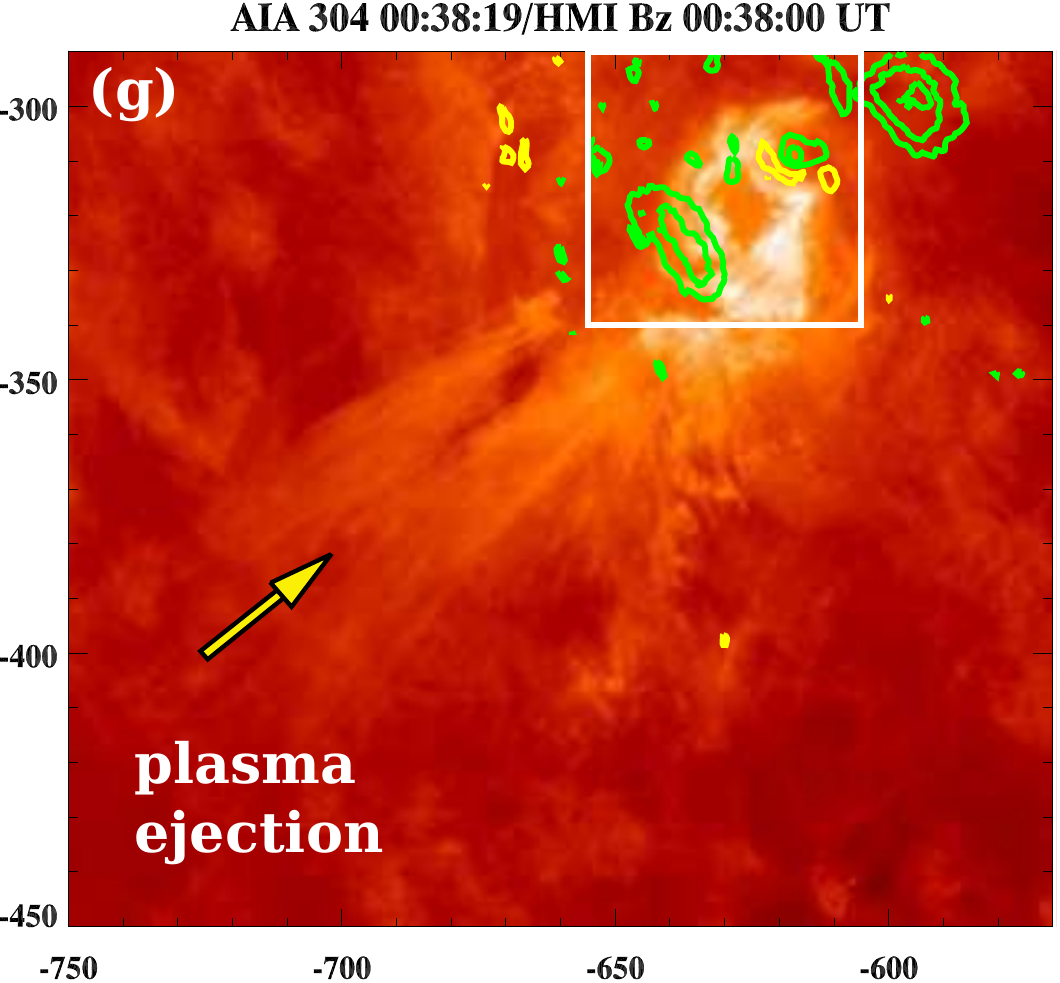}
\includegraphics[width=5.3cm]{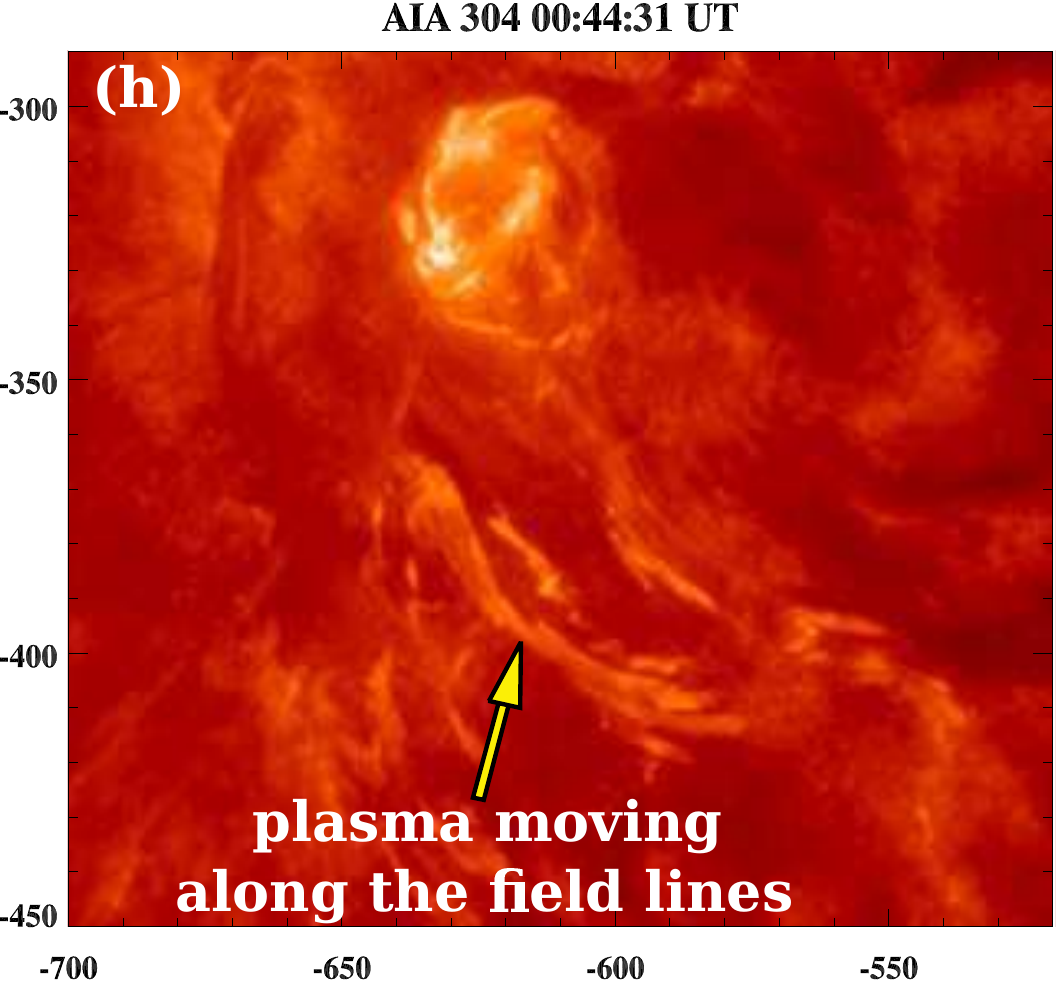}
\includegraphics[width=5.3cm]{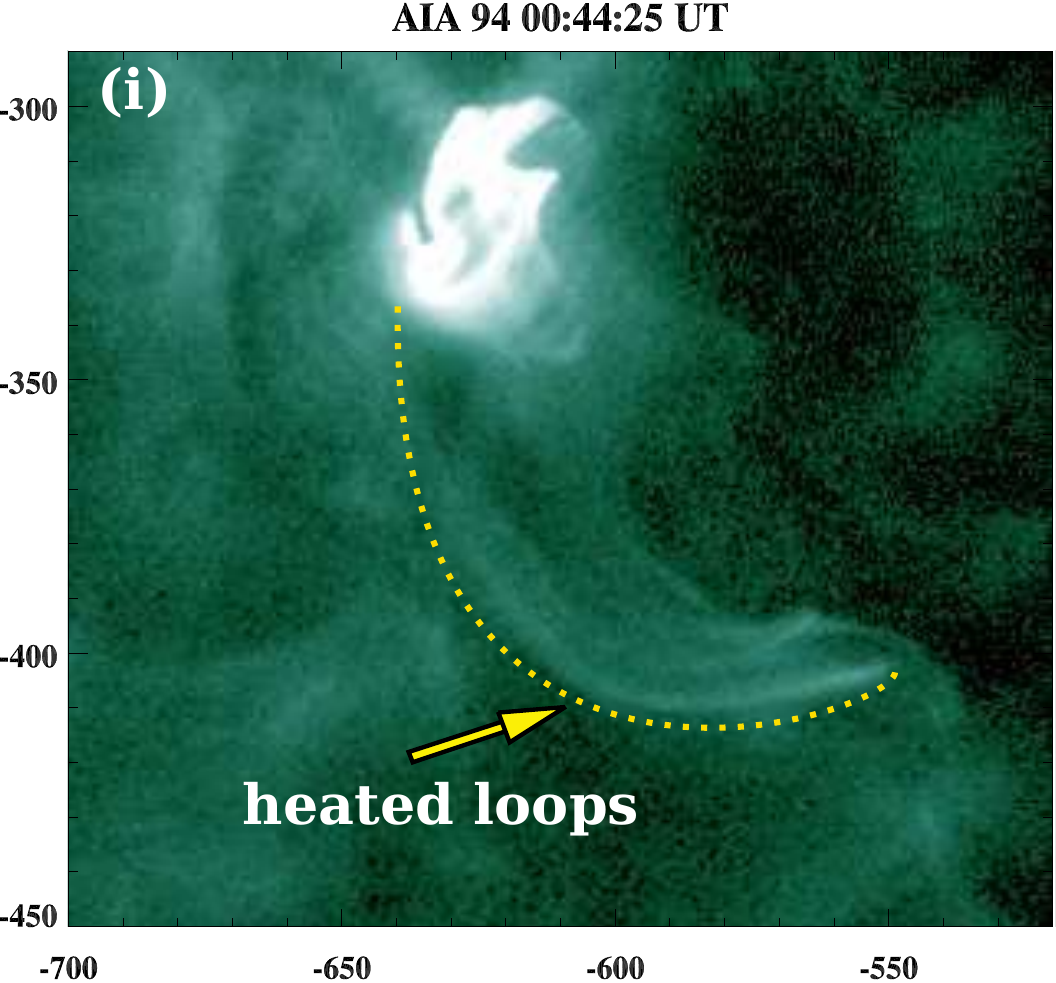}

}
\caption{(a-c) IRIS 1300 \AA~ images (T$\sim$20,000 K) showing the eruption of the flux rope on 13 June 2014. (d-f) AIA 1600, 171 and 131 \AA~ images of the flux rope eruption. The first image of each panel is overlaid by HMI magnetogram contours of positive (green) and negative (yellow) polarities. (g-i) AIA 304 and 94 \AA~ images of the surge like ejection and plasma moving along the field lines connected to the flare site.  (An animations of this figure is available online)}
\label{flare2}
\end{figure*}

The flux rope erupted during the second C8.5 flare that occurred after 3 hours of the first M1.0 flare. Figure \ref{flare2}(a-c) displays IRIS 1300 \AA~ (C II, T=20,000 K) slit-jaw images of this C8.5 flare. Figure \ref{flare2}(a) is overlaid with HMI magnetogram contours (positive (green) and negative (yellow) polarities) to show the position of the flux rope and identification of its footpoints. The flare brightening starts below the flux rope at the same location between P1 and N1. One may observe the rise of the flux rope with enhancement of the flare brightening at 00:34 UT. The footpoints (F1 and F2) of the twisted flux rope are clearly identified. Figure \ref{flare2}(d-f) shows AIA 1600, 171 and 131 \AA~ images of the flare site. The flux rope was observed in these channels also. AIA 131 \AA~ image shows the initial flare brightening near footpoint F1 of the flux rope. This is the same site where the onset of the first M1.0 flare occurred.
The erupting flux rope most likely interacted with the overlying fields. We observed surge like cool plasma ejections with untwisting motion (see AIA 304 \AA~ movie).  Bottom panels (g-i) of Figure \ref{flare2} show AIA 304 and 94 \AA~ images. We noticed the plasma ejection radially outward and horizontally along the heated loops (AIA 94 \AA). The AIA 94 \AA~ image shows that one footpoint of the heated loop systems connected to the flare site.

\begin{figure*}
\centering{
\includegraphics[width=6.0cm]{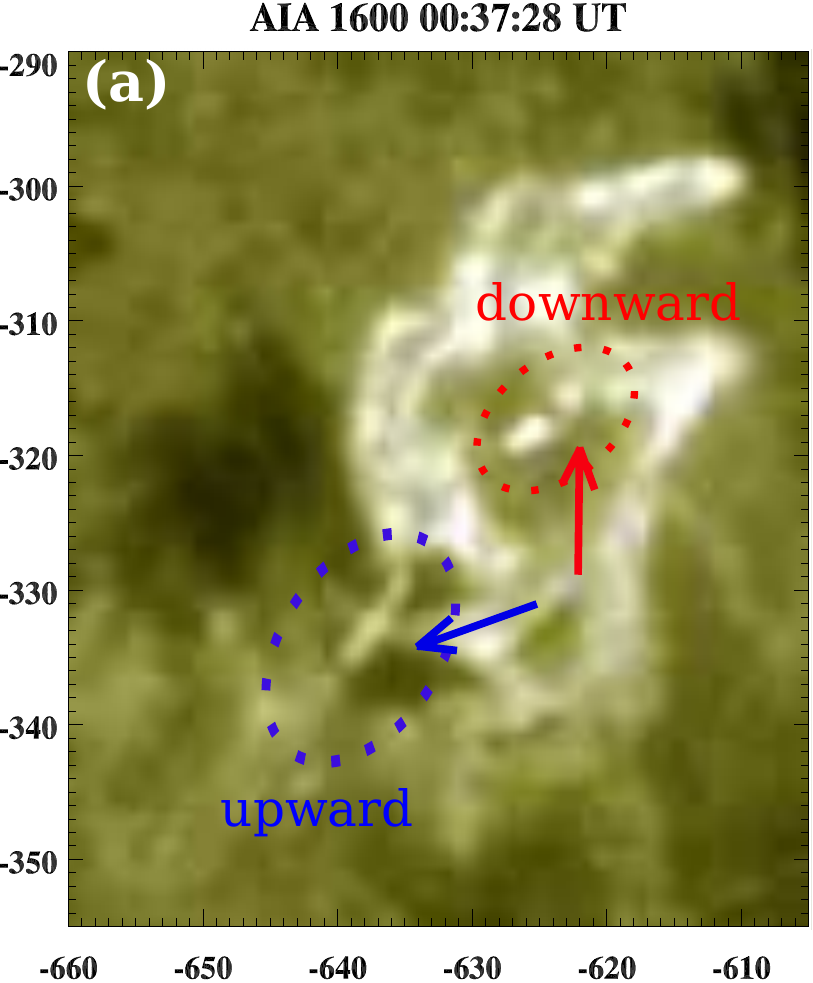}
\includegraphics[width=6.0cm]{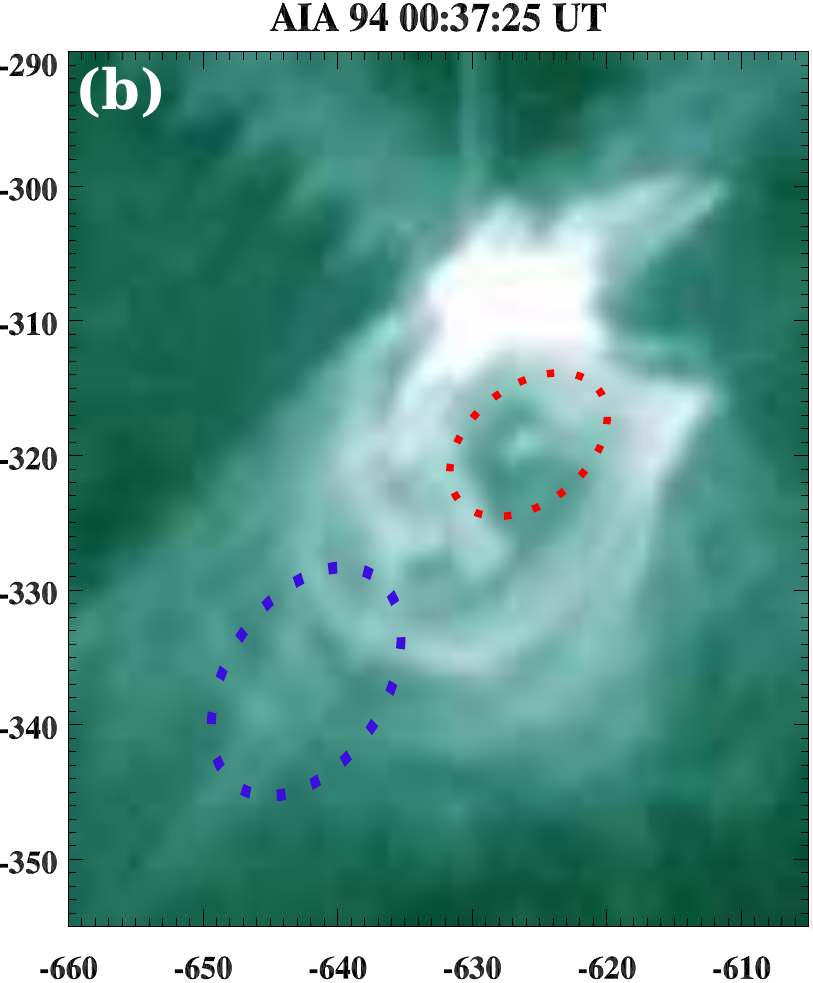}

\includegraphics[width=8.0cm]{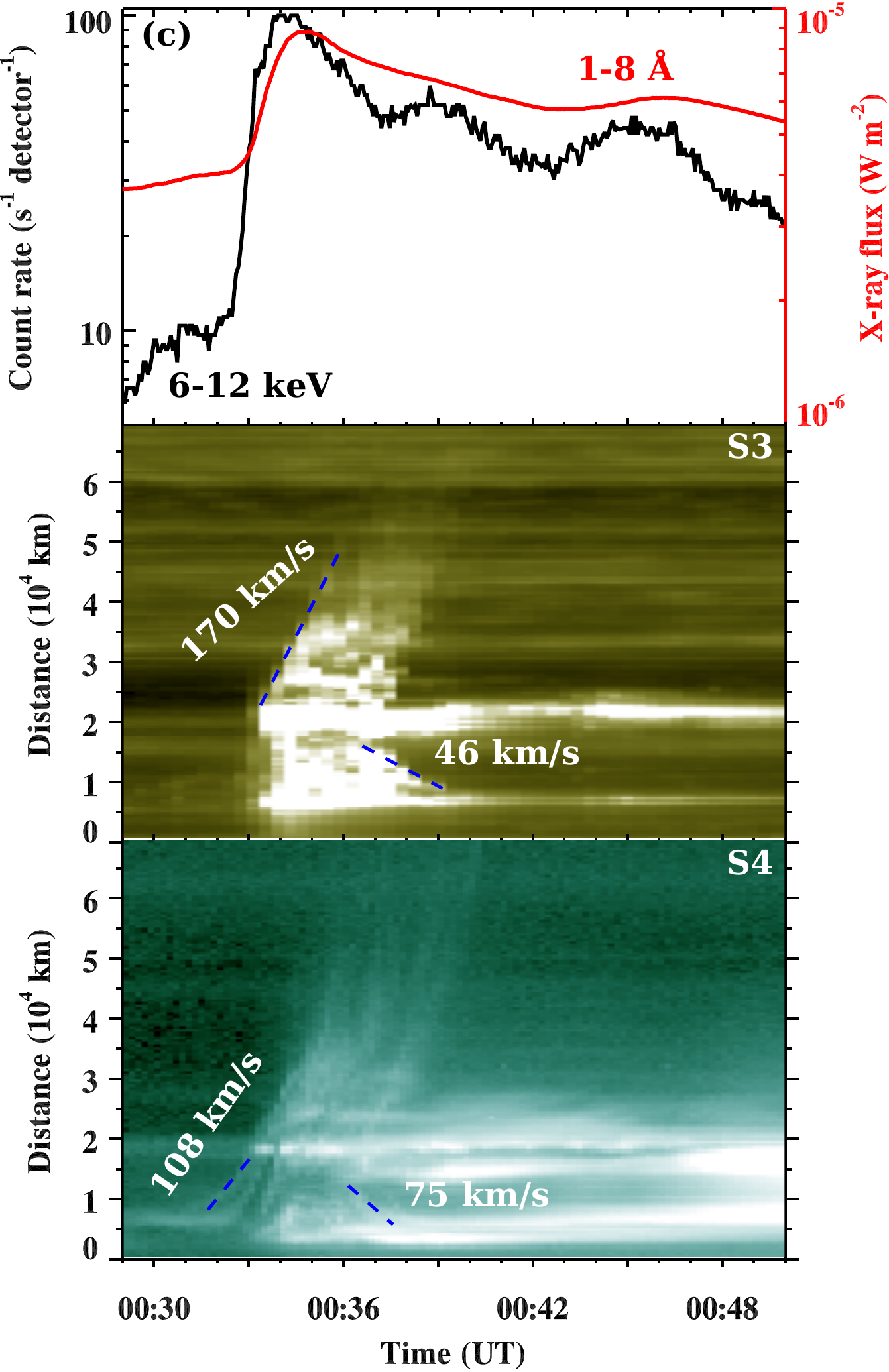}
\includegraphics[width=5.5cm]{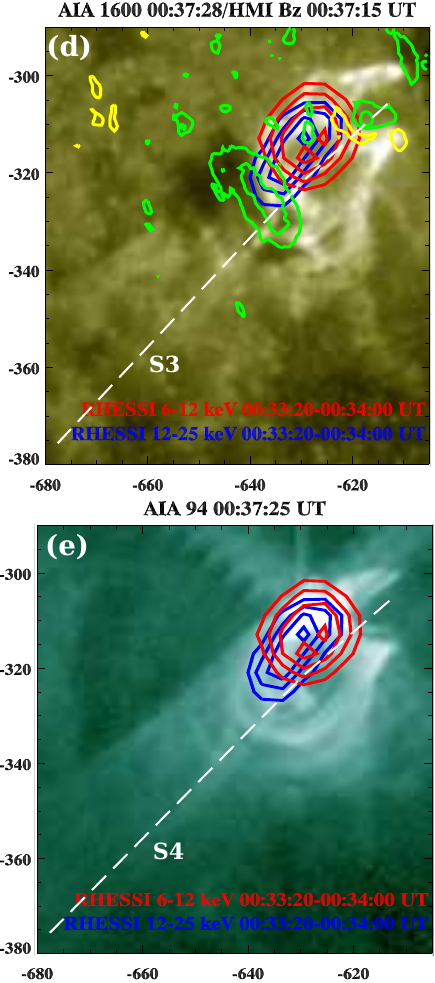}
}
\caption{(a-b) AIA 1600 and 94 \AA~ images of the flare site showing the upward (blue) and downward (red) moving plasma. (c) RHESSI X-ray flux in 6-12 keV channel plotted with GOES soft X-ray flux (1-8 \AA, red) for the second C8.5 flare.  Stack plots along slices S3 and S4 shown in AIA 1600 and 94 \AA~ channels. (d-e) AIA 1600  and 94 \AA~ images overlaid by RHESSI X-ray contours in 6-12 keV (red) and 12-25 keV (blue). AIA 1600 \AA~ image is overlaid by HMI magnetogram contours of positive (green) and negative (yellow) polarities. (An animation of this figure is available online)}
\label{stack}
\end{figure*}

To investigate details of the flux rope eruption and its interaction with the ambient field, we used AIA 1600 and 94 \AA~ images. Figure \ref{stack}(a,b) show AIA 1600 and 94 \AA~ images during the flare maximum phase ($\sim$00:37 UT). AIA 1600 and 94 \AA~ movies show the ejection of an untwisting surge as a result of the flux rope interaction with the overlying fields. We also noticed bidirectional plasma injection. AIA 1600 \AA~ image shows the hot plasma blobs moving downward (red dotted ellipse) and upward (blue dotted ellipse). A similar plasma blob (downward) was observed in the AIA 94 \AA~ images.
To track the motion of these plasma ejections, we chose slice cuts S3 and S4 in AIA 1600 and 94 \AA~ images (Figure \ref{stack}(c,d)). Figure \ref{stack}(c) shows distance-time plot of the AIA 1600 (middle) and 94 (bottom) \AA~ intensity along the selected slices S3 and S4. The top panel displays GOES soft X-ray flux (1-8 \AA) profile (red) along with RHESSI X-ray flux in 6-12 keV energy band. Interestingly, we observed quasi-periodic pulsations (QPP, period$\sim$5 min) in the 6-12 keV channel, which are quite similar to recently reported by \citet{kumar2015}. Initially, the flux rope rose with the speed of $\sim$108 km s$^{-1}$. The speed of the surge like ejection was $\sim$170 km s$^{-1}$. The hot plasma blob downflows with a speed of $\sim$46-75 km s$^{-1}$. Figure \ref{stack}(d,e) display RHESSI X-ray contours in the 6-12 (red) and 12-25 (blue) keV, which are overlaid on AIA 1600 and 94 \AA~ images. Contours of an HMI magnetogram of positive (green) and negative (yellow) polarity are overlaid on the AIA 1600 \AA~ image. The location of the X-ray sources is almost similar to that observed for the first M1.0 flare. The hard X-ray sources are formed above the center of the quasi-circular ribbon. These sources most likely show the particle acceleration site in the corona (i.e., loop-top sources). The interaction of the flux rope with the overlying arcade loops can trigger a magnetic reconnection and associated plasma heating. Magnetic reconnection above the flux rope can cause the formation of the hard X-ray sources. Recently, \citet{kumar2014} observed formation of X-ray sources above a kink unstable filament, which was failed to erupt. The X-ray sources above the filament or flux rope suggest the particle acceleration site above the filament as a result of magnetic reconnection. In our case, the existence of surge like multiple ejections of upward and downward moving plasma blobs is indirect evidence of magnetic reconnection of an erupting flux rope with the ambient coronal fields.

\begin{figure*}
\centering{
\includegraphics[width=4.5cm]{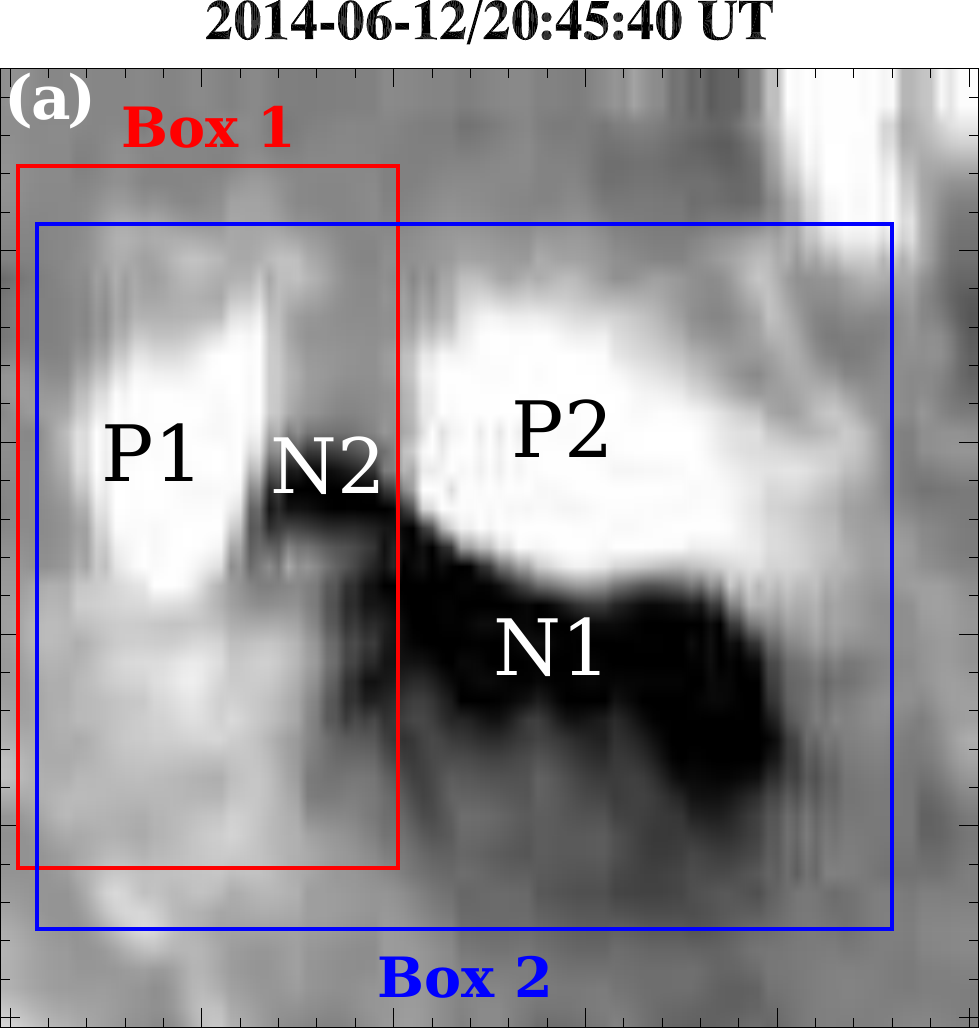}
\includegraphics[width=4.5cm]{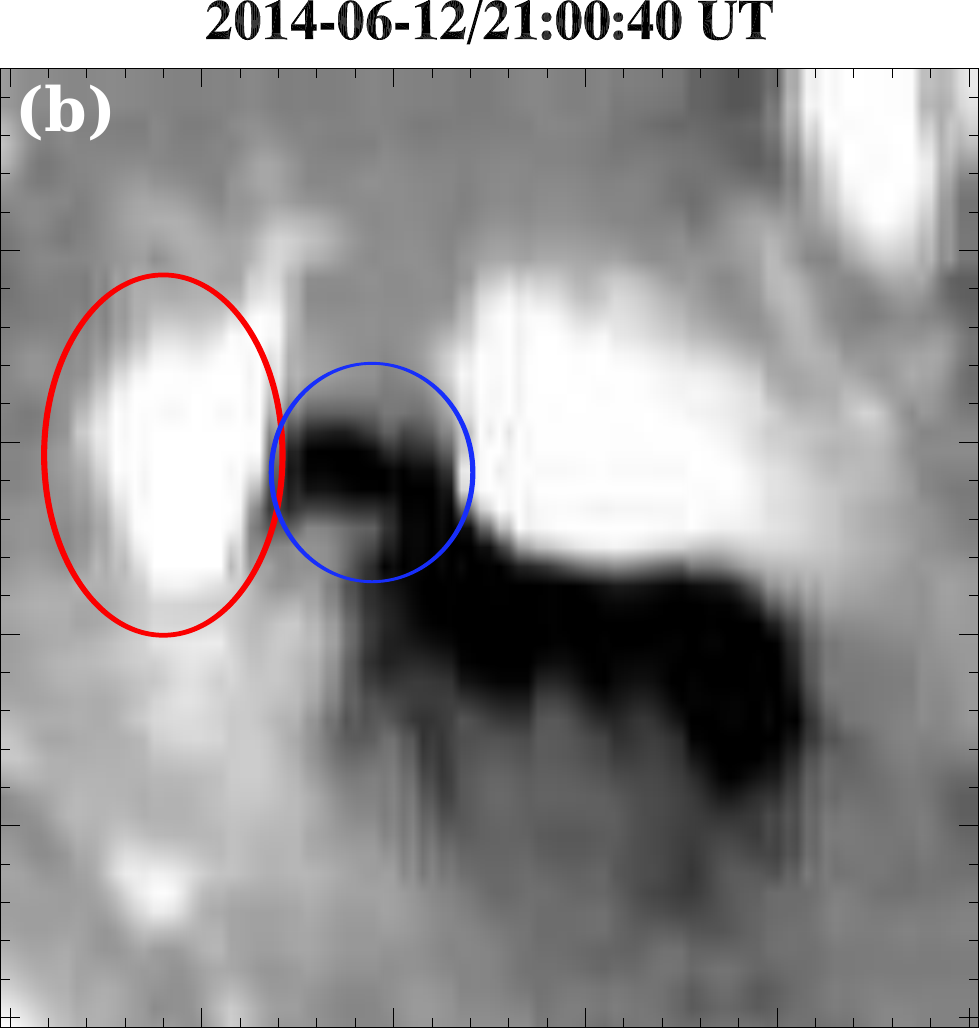}
\includegraphics[width=4.5cm]{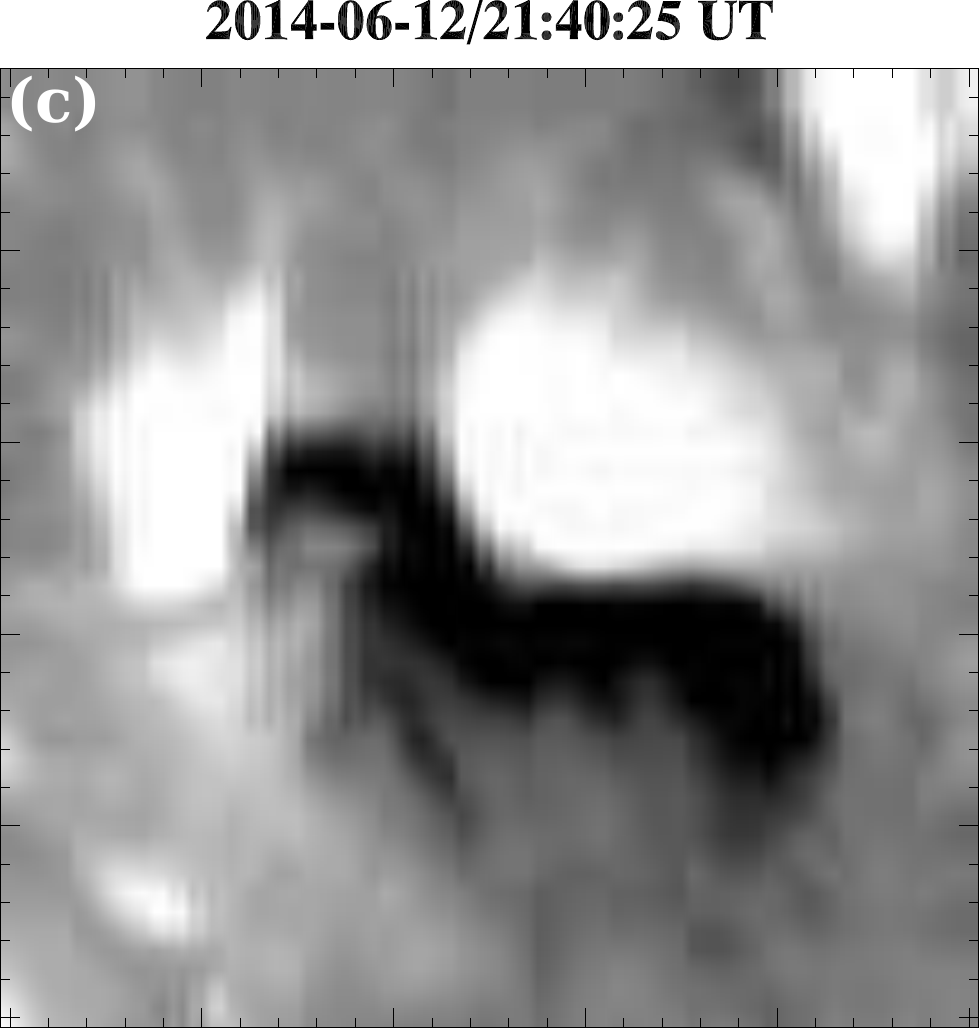}

\includegraphics[width=4.5cm]{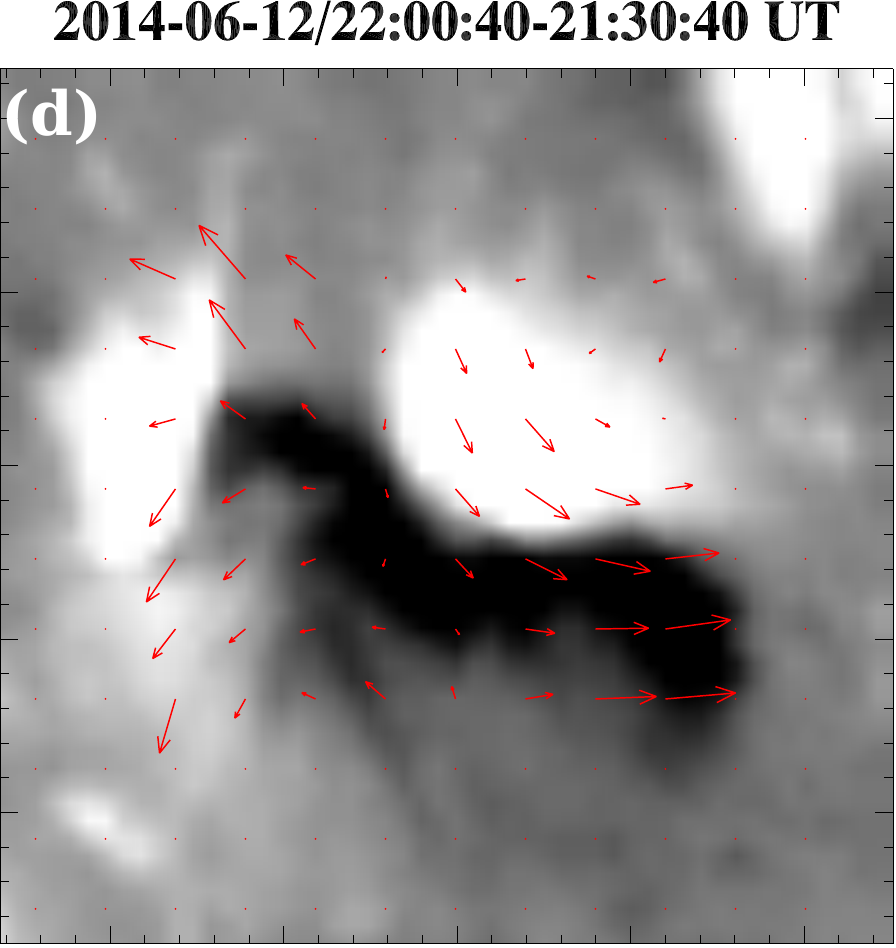}
\includegraphics[width=4.5cm]{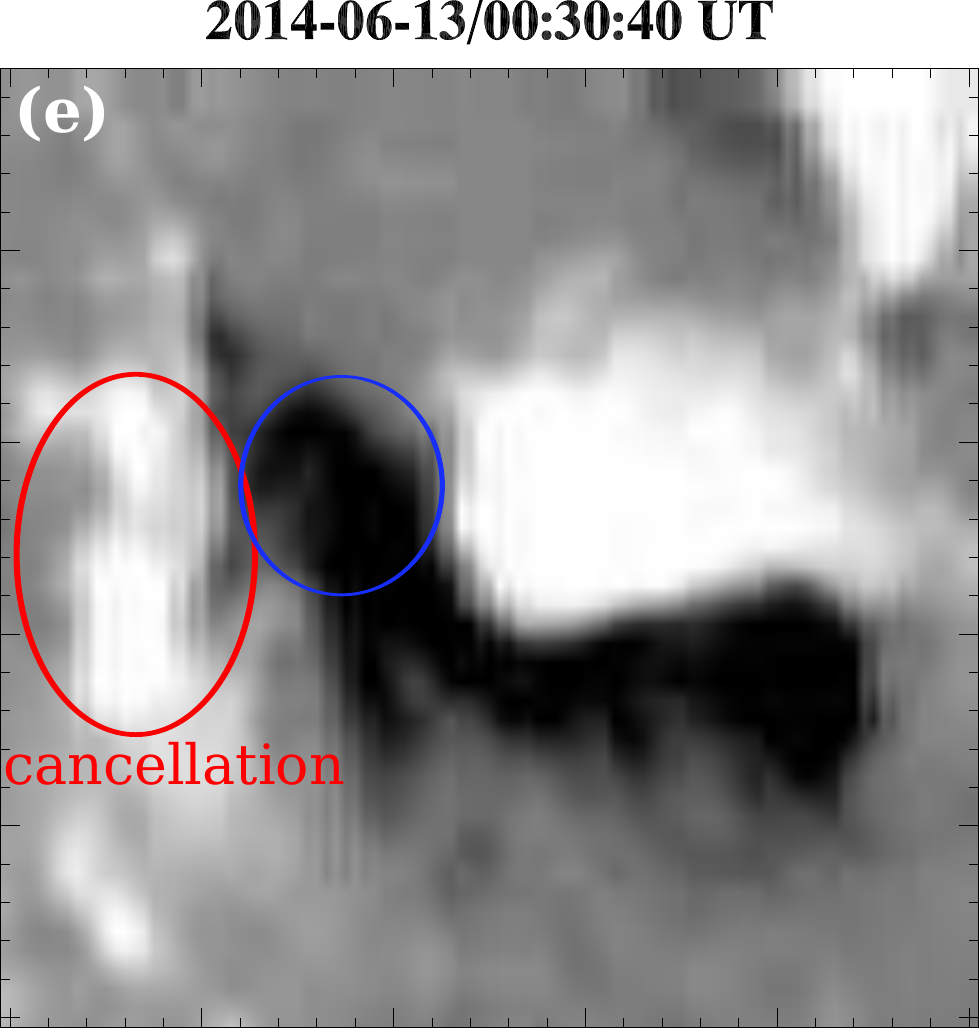}
\includegraphics[width=4.5cm]{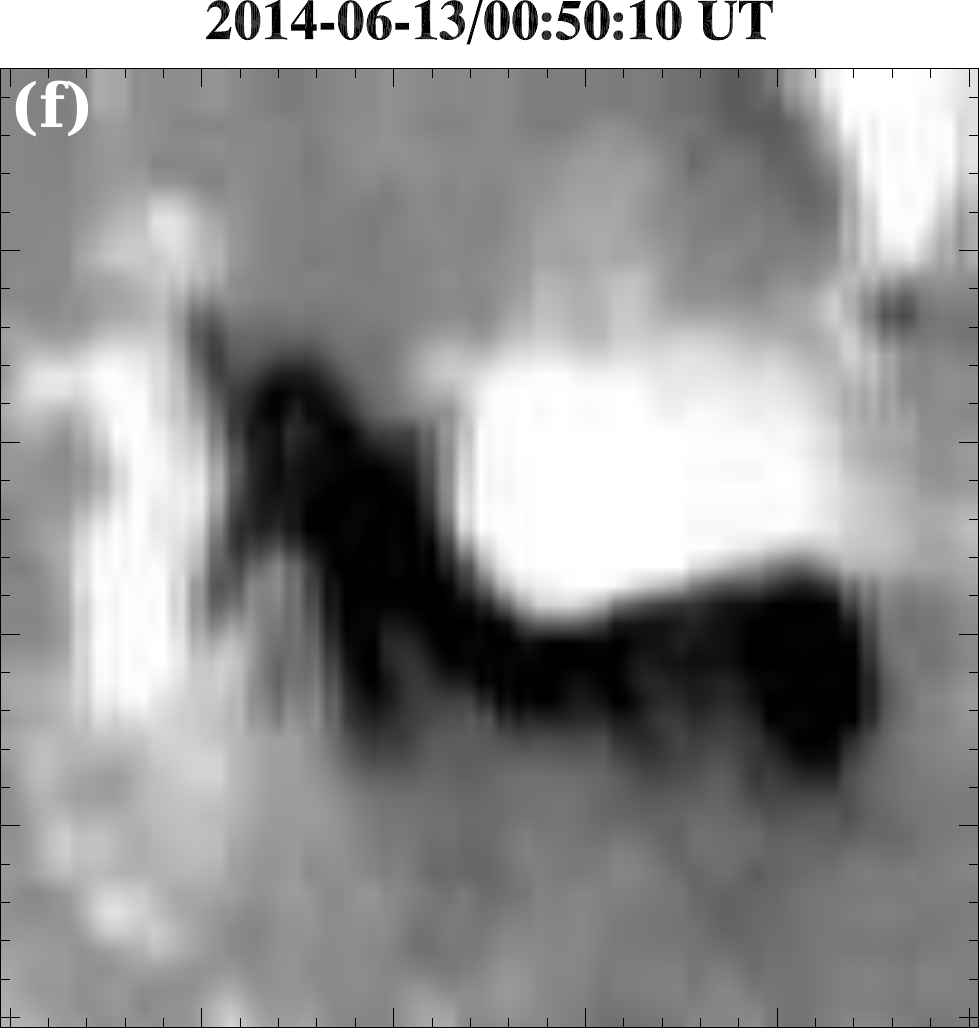}

\includegraphics[width=8.5cm]{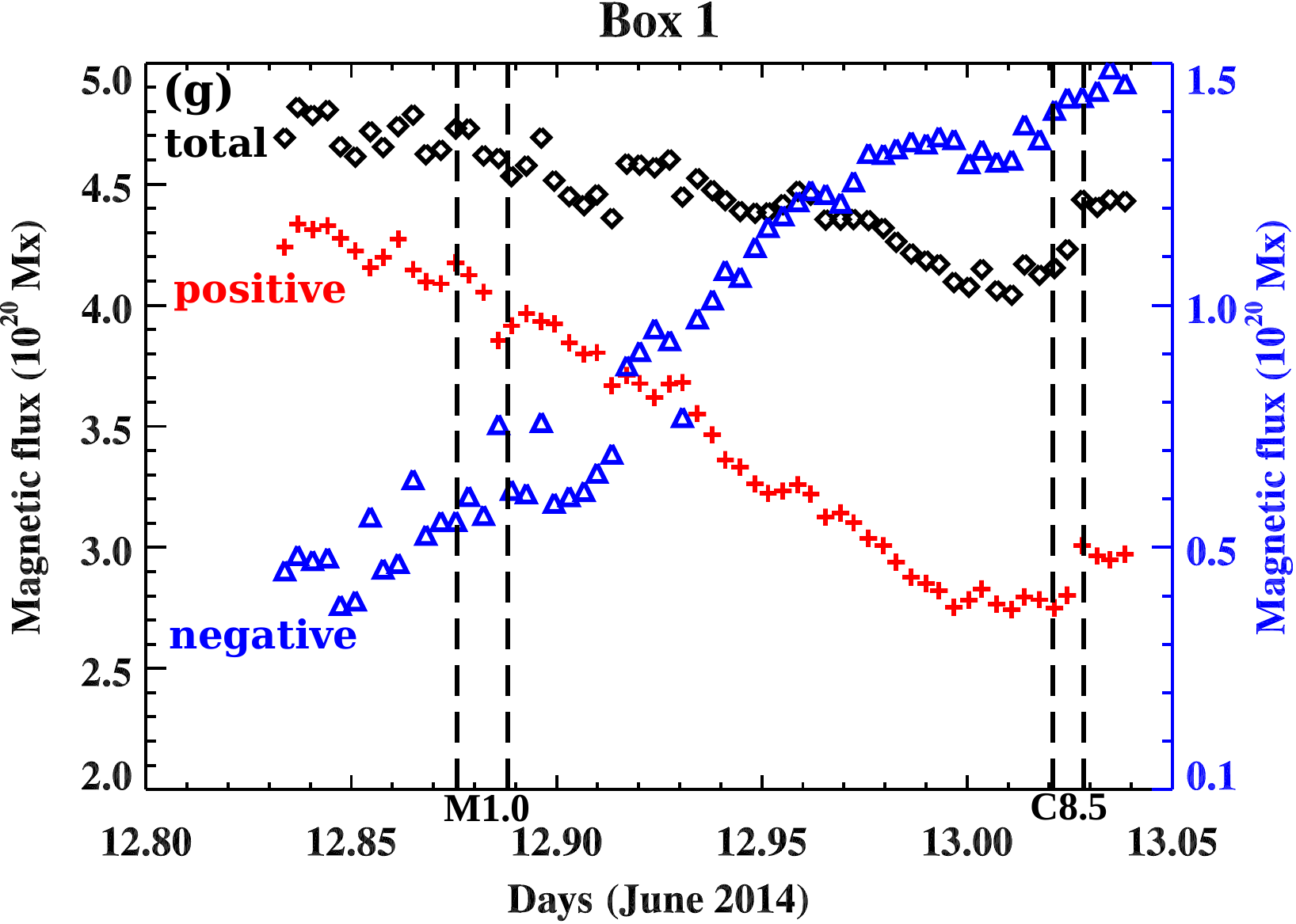}
\includegraphics[width=8.5cm]{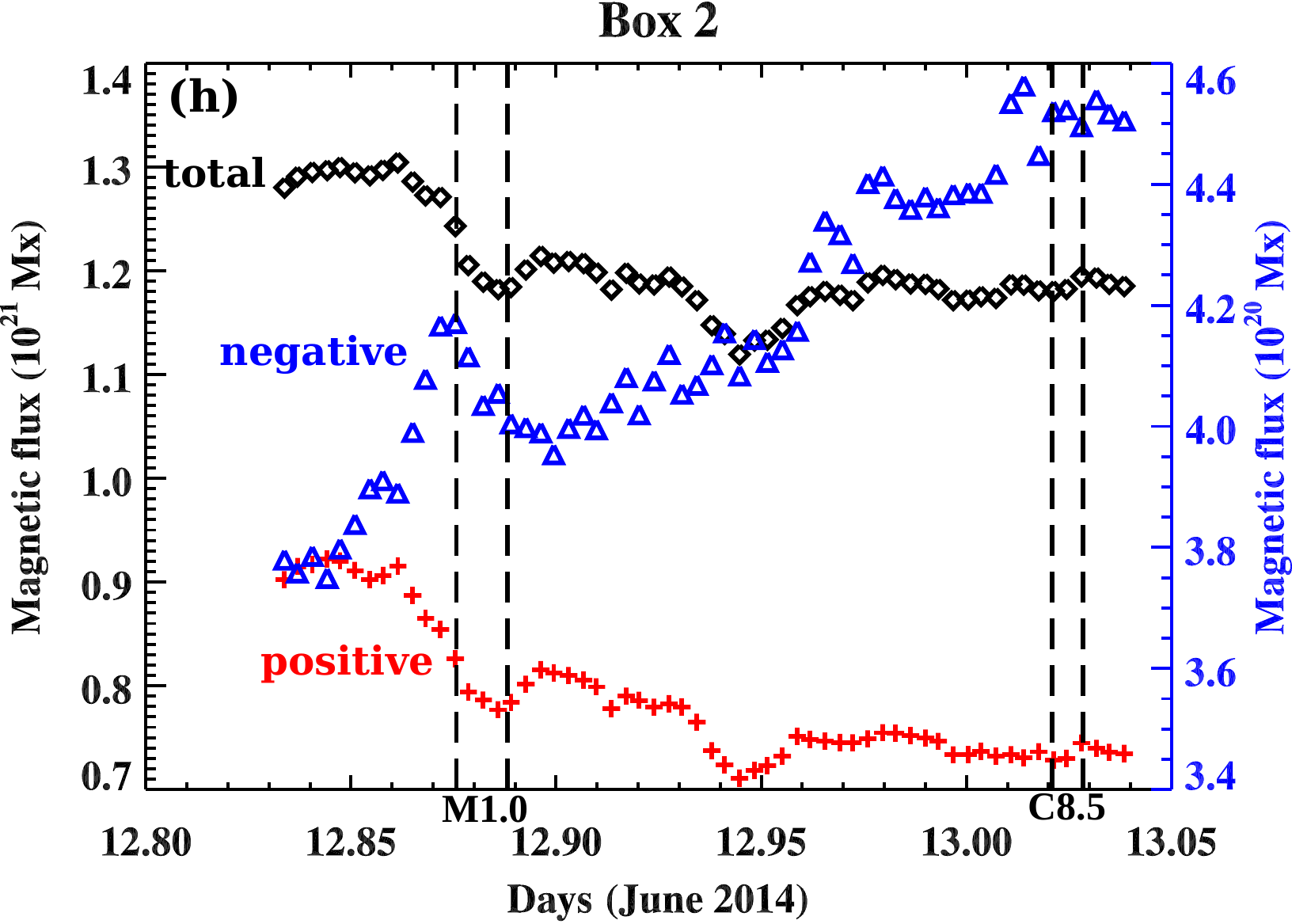}
}
\caption{(a-f) Selected HMI magnetograms showing the evolution of magnetic fields during the flares. The red and blue ellipses indicate changes in the positive and negative polarity sunspots. (g-h) Positive (red), absolute negative (blue) and total (black) flux profiles within box 1 and 2. Two vertical dotted lines represent the start and end times of the two flares (M1.0 and C8.5). (An animation of this figure is available online)}
\label{hmi}
\end{figure*}


\begin{figure*}
\centering{
\includegraphics[width=8cm]{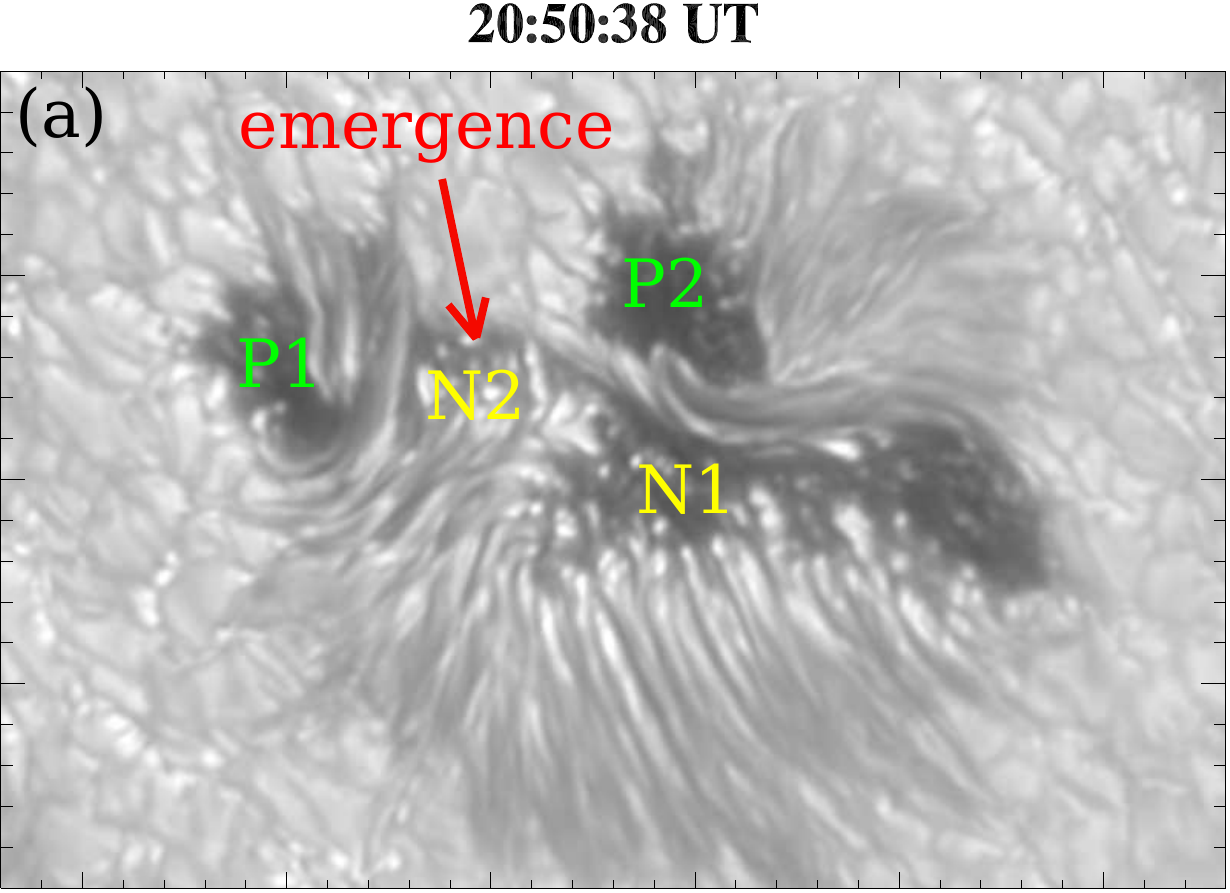}
\includegraphics[width=8cm]{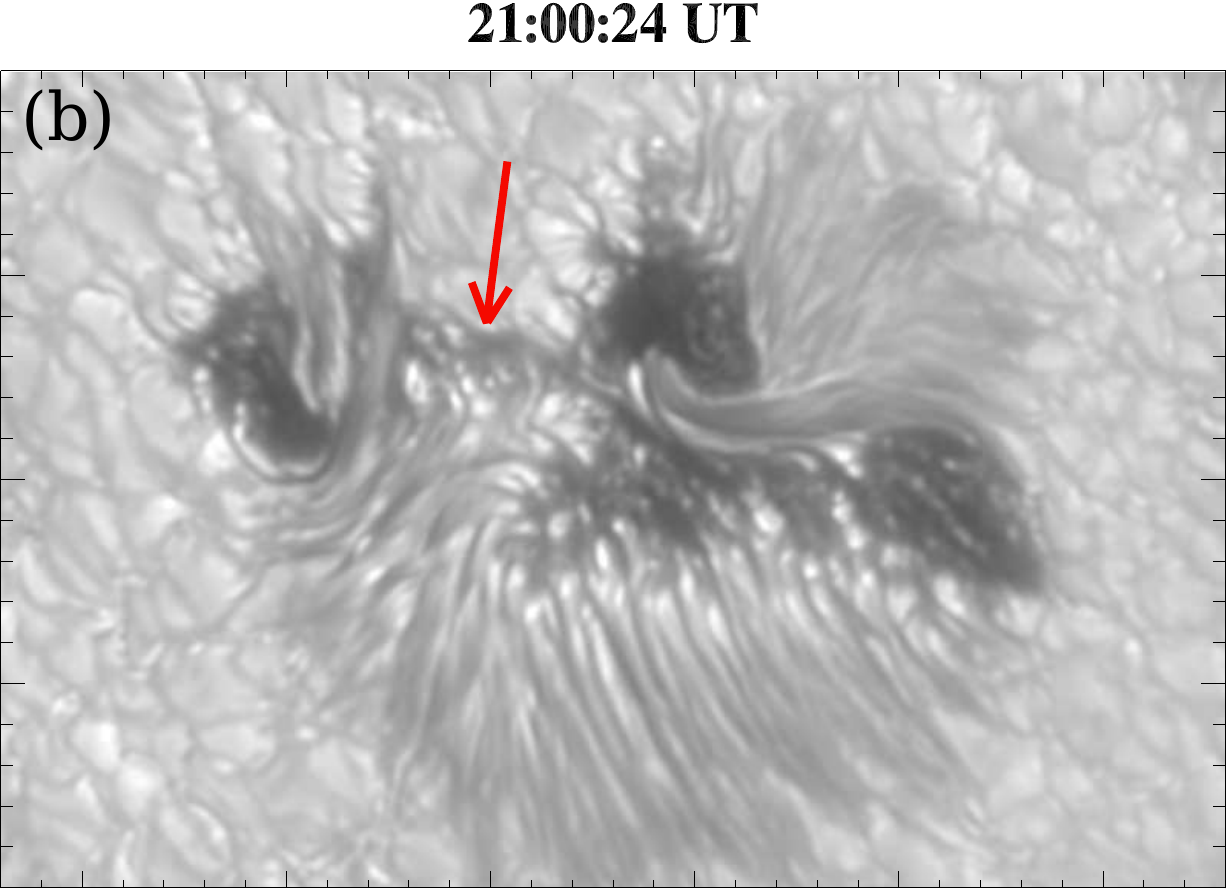}

\includegraphics[width=8cm]{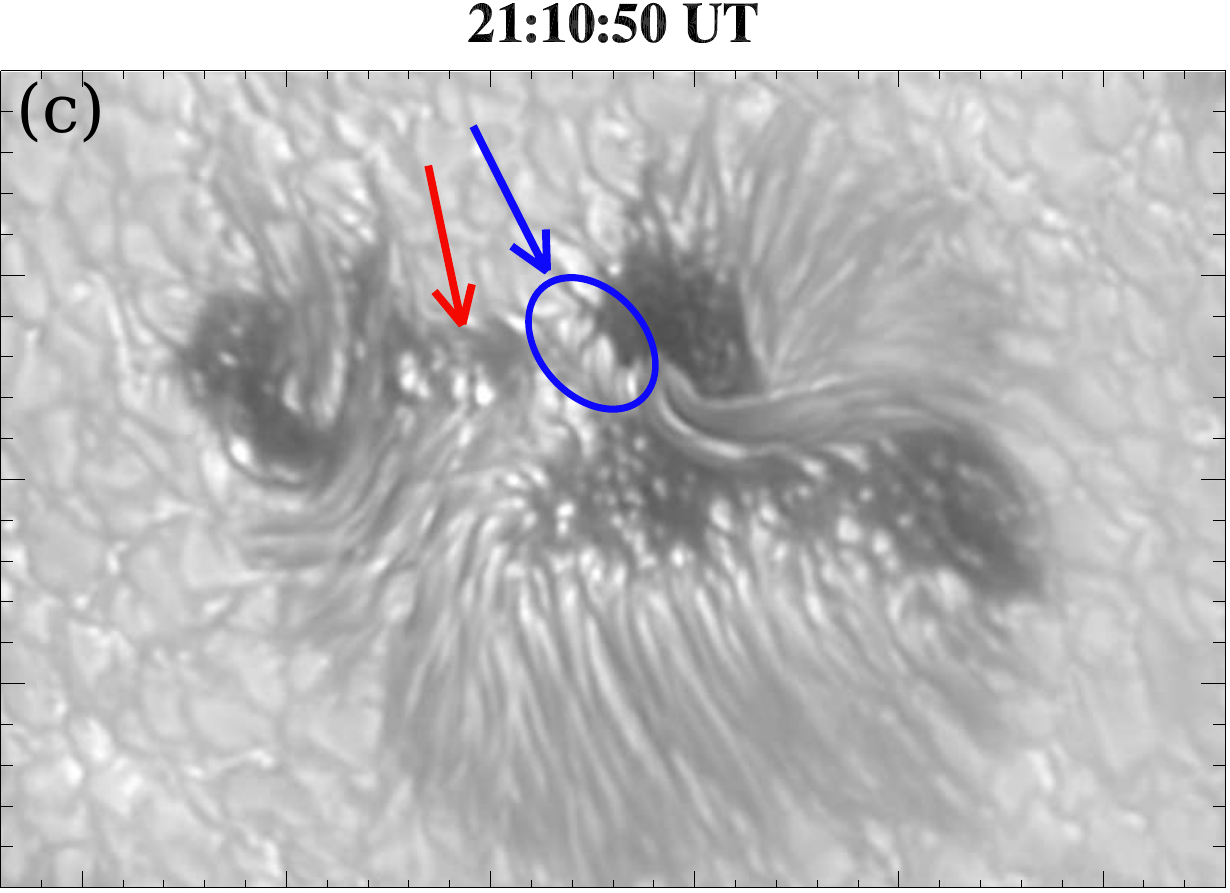}
\includegraphics[width=8cm]{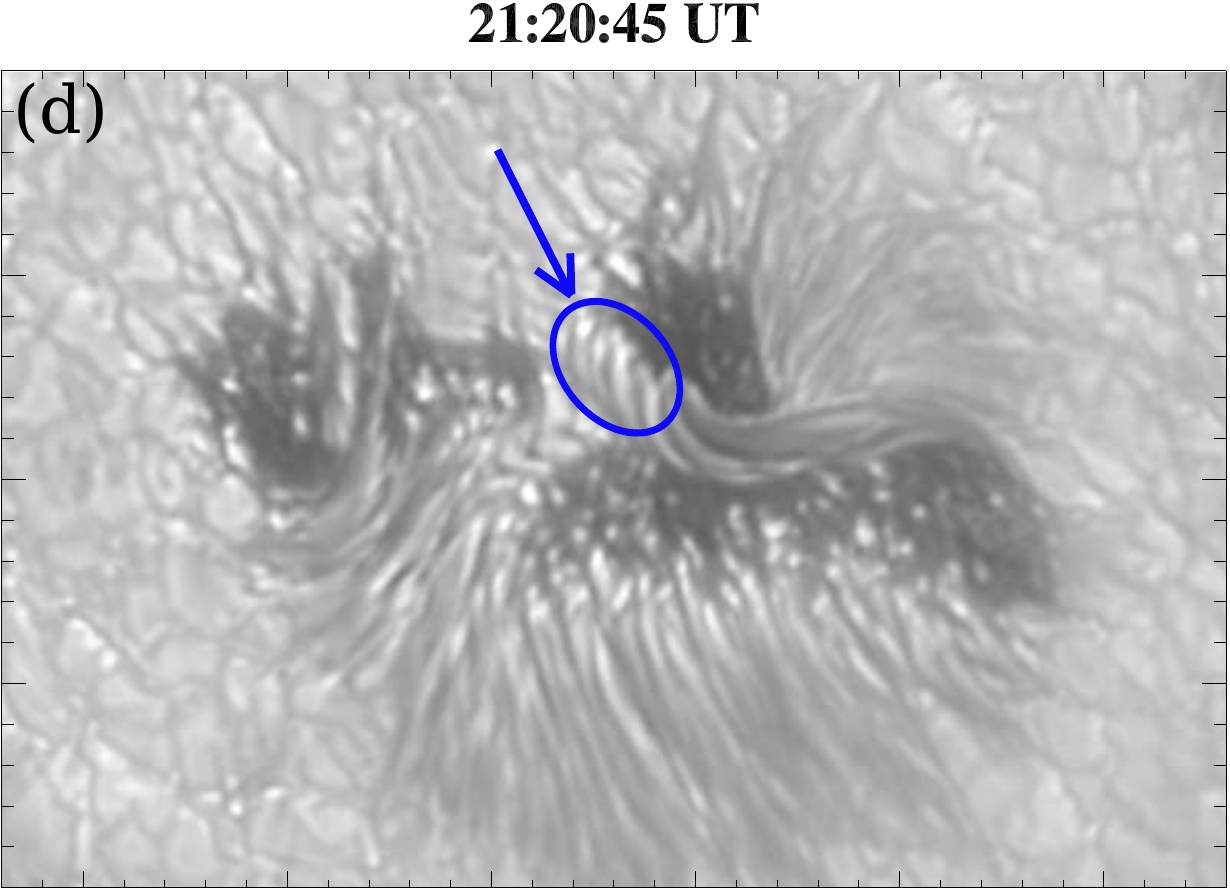}
}
\caption{TiO image sequence showing new flux emergence during the first M1.0 flare. Red arrow indicates the pore like emerging flux (negative), whereas blue one shows the expansion of the flux tube anchored in the positive polarity sunspot. (An animation of this figure is available online)}
\label{tio}
\end{figure*}


\subsection{Evolution of magnetic field}
To understand the changes in the photospheric magnetic fields associated with the flare and the eruption, we used HMI magnetograms acquired before, during and after the two flares studied here (Figure \ref{hmi}). An HMI magnetogram movie shows the continuous flux emergence of negative polarity region between P1 and P2, and cancellation of positive polarity field P1. In addition, N1 (between P1 and P2) continuously displaces P1, and it moves down (toward south) associated with flux cancellation. Figure \ref{hmi}(a,b) display magnetogarms taken before the first M1.0 flare. Note that flare and small jets erupt between P1 and N1 (see, Figure \ref{hab4}(b)). We noticed an increase in the negative flux before the flare onset (21:00 UT, marked by the blue ellipse). Figure \ref{hmi}(d) shows a photospheric flow map derived from the differential affine velocity estimator (DAVE) method \citep{schuck2006}. We chose 30 min time-difference between two selected magnetograms, and a window size of 10$\arcsec$. The flow map at 22:40 UT shows  converging and shearing flows also evident in the HMI magnetogram movie. The longest arrow corresponds to the flow speed of 210 m s$^{-1}$. The direction of arrows suggest that the negative polarity element (N1) pushes P1, and P1 moves down. The negative flux emergence and shear flows helped in the build-up of magnetic energy.

To estimate the amount of the emerged and canceled flux, we selected the boxes 1 and 2 in Figure \ref{hmi}(a). The evolution of positive (red), absolute negative (blue) and total (black) magnetic flux (in Mx) is shown in Figure \ref{hmi}(g,h) within the selected boxes 1 and 2, respectively. The start time and end time of the both flares are marked by two vertical dotted lines. Before, the first M-class flare, we see continuous emergence of the negative flux whereas cancellation of positive flux in both boxes. The emergence of negative flux is mainly responsible for the first M1.0 flare. Interestingly, in box 2 we observed flux cancellation (both positive and negative) during the first flare, which results in an decrease in   the total flux. Similar behavior is also seen in box 1, but not so clearly. This is very important because it supports our interpretation that the flux rope was formed during the first M1.0 flare as a result of flux cancellation. After the first flare, the emergence of negative flux continues before the start of the second flare whereas the positive flux decreased. However, we do not see significant changes in positive and negative fluxes during the second C8.5 flare. The flux rope erupted during the second C8.5 flare.

To check the evolution of fine photospheric structures (at the flare site) associated with the M1.0 flare, we used TiO images from NST. Figure \ref{tio} shows the sequence of high-resolution TiO images before and during the M1.0 flare. We see a growing pore like region (marked by red arrow) during the M1.0 flare. This is emerging negative polarity region that continuously pushes P1. Untwisting jets (in the chromosphere) observed before the M1.0 flare emanate above this growing spot (see the TiO movie). Another interesting feature that we note is an emergence of a twisted flux tube at the edge of P2 (marked by blue arrow). This twisted feature starts developing during the flare maximum at 21:10 UT and grows continuously later. This expanding flux tube has some connection with the formation of the flux rope (for details refer to the discussion part). 

\begin{figure*}
\centering{
\includegraphics[width=8.9cm]{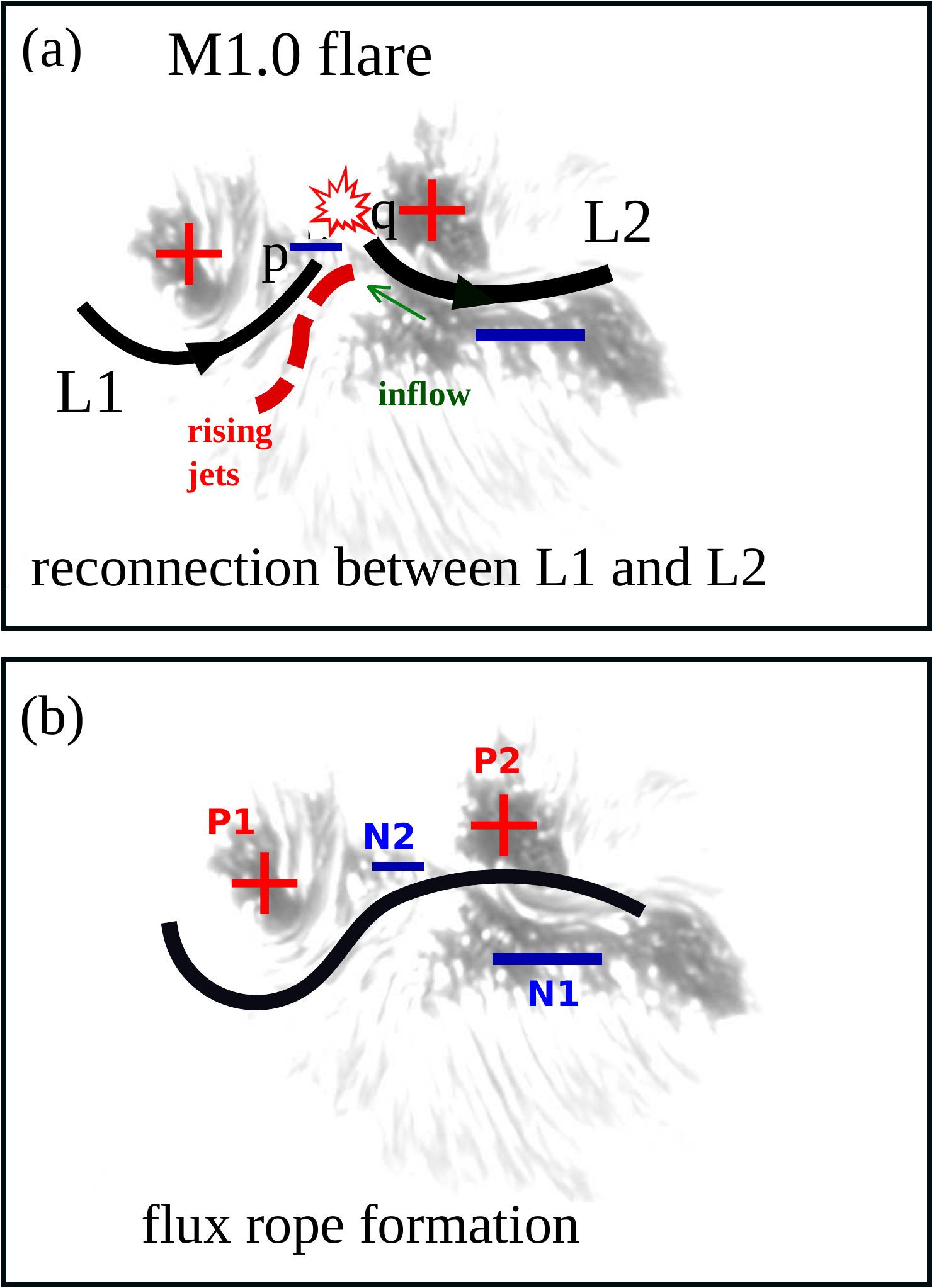}
\includegraphics[width=6cm]{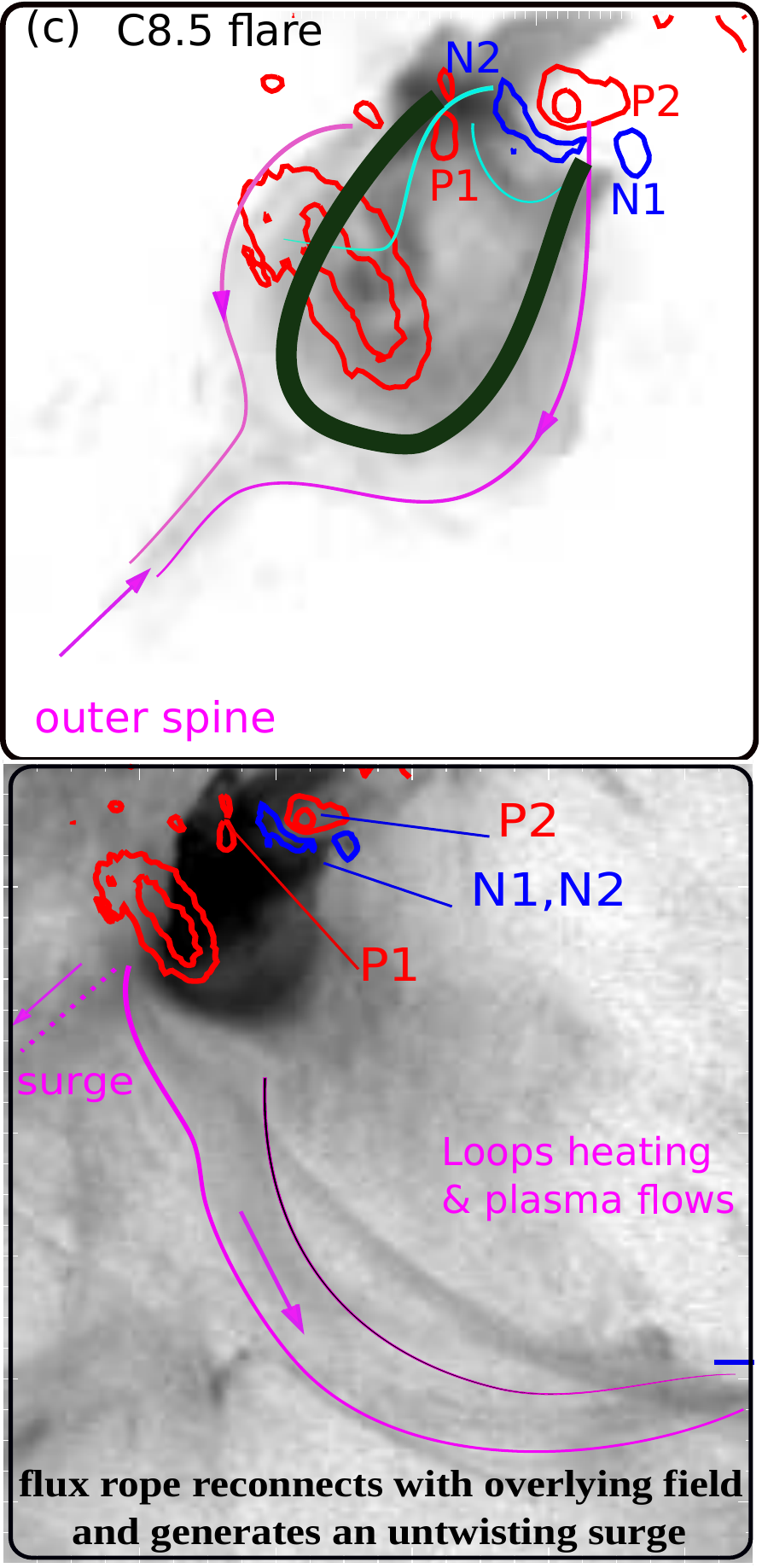}
}
\caption{(a-b) Schematic cartoon (over the NST TiO image at 20:50:38 UT) showing the formation of the flux rope by reconnection between L1 and L2 loops during the M1.0 flare. P1, P2, N1, and N2 are small sunspots of positive and negative polarities. (c) Flux rope reconnection with the ambient fields and formation of the hot arcade loops during the C8.5 flare. The possible connectivity of the field lines is drawn over the AIA 131 \AA~ image during the second flare. The red and blue contours denote positive and negative polarities. N1, N2 are surrounded by positive polarities (three sides), therefore, forms a quasi-circular PIL and flare ribbons. The cyan field lines show the low lying loops connecting to the opposite polarity sunspots. The purple lines are the field lines emanating from the positive polarity spots reaching to the outer spine and connecting to the remote negative polarity in the southward direction. These loops (purple) were heated during the magnetic reconnection and showed plasma flows along them. The red arrows indicate the direction of the untwisting surge along and across the heated loops.}
\label{cartoon}
\end{figure*}


\section{SUMMARY AND DISCUSSION}
 We presented high-resolution observations of the small flux rope formation in the chromosphere during an M1.0 flare and its eruption during the C8.5 flare. The twisted emerging fluxes from opposite polarity sunspots (P1 and N1) reconnected (as a result of tiny negative flux emergence at the neutral line) to form a twisted flux rope in the chromosphere. The direction of the twisted threads suggests the formation of a right handed S shaped flux rope. Rising small jets ($\sim$20-55 km s$^{-1}$) with untwisting motion were observed before the trigger of the first M1.0 flare. Strong plasma inflow ($\sim$10 km s$^{-1}$) was seen below the jet (between two sheared H$\alpha$ loops) after its escape. The evacuation below the rising jet possibly trigger the cool plasma inflow in the chromosphere. The surrounding plasma below the untwisting jet moved toward the PIL. This mechanism is quite similar to the plasma inflows observed (in the corona) during the ejection of a rising large-scale flux-rope in the corona \citep{chen2000,savage2012,kumar2013b}. A similar unidirectional inflow structure was recently reported by \citet{mulay2014} in the AIA 304 \AA~ channel that triggered magnetic reconnection below a rising flux rope. However, this type of cool plasma inflow feature in such a high resolution is not reported before.
The ribbon separation (speed$\sim$1.5-1.6 km s$^{-1}$) progresses after the ejection of the small jet which is consistent with a typical two-ribbon flare associated with a filament eruption. However, this speed is much smaller in comparision to a typical two ribbon flares. In our case, ribbon separation (within global quasi-circular ribbon) was associated with the eruption of the untwisting jet.

\citet{kumar2010} observed the trigger of a M-class flare caused by the interaction of two filament channels approaching to each other with a speed of $\sim$10 km s$^{-1}$. Inflow speed observed in our case is consistent with the filaments approaching speed in \citet{kumar2010}.
Our observations suggest that surge like ejection during the second flare is associated with the destruction of the small flux rope by reconnection with the ambient overlying fields. However, this type of small flux rope was not observed in such details in the earlier low resolution H$\alpha$ observations before the NST and IRIS.

Using IRIS spectral data, \citet{sadykov2014} studied the first M1.0 flare and reported strong red-shifts (during untwisting jets) and blue shifts (chrmospheric evaporation) during the pre-flare and decay phase respectively. The motions in the untwisting jet are toward the south-east direction. The position of the IRIS slit was across the jet. The strong redshifts ($\sim$100 km s$^{-1}$), observed during the preflare phase, was most likely associated with the downflows associated with the reconnection driving the untwisting jets. Recently, \citet{tian2014} also reported strong redshifts ($\sim$125 km s$^{-1}$) using the IRIS data and interpreted them as being a reconnection generated downflow/hot retracting loops.
It is likely that slow reconnection starts (20:50 UT onwards) during the pre-flare heating phase generating small untwisting jets, which probably follow the inner spine of the fan loops. Later, we noticed the rotation of the field lines around the spine in the AIA 94 \AA~ movie during the first flare, which supports the 3D torsional spine reconnection \citep{pontin2007,priest2009}. 

\citet{kumar2013blob} observed rising of a untwisting cool plasma blob that initially generated a two-ribbon (R1 and R2) flare, and formed another third ribbon (R3) while reaching a certain height. Initial ribbon separation (R1 and R2) was associated with the rise of the untwisting blob/jet. The mechanism of the ribbon separation in their case was quite similar as observed in the filament eruption. In addition, the global morphology of the flare ribbons during the flare maximum was circular, and the magnetic field configuration had a fan-spine topology (please, refer to Figure 1 of their paper). Using NST data, \citet{wang2014} also observed three ribbons flares in a fan-spine topology associated with jets.  

During both flares hard X-ray sources (12-25 keV) are formed over the center of the quasi-circular ribbon, which is most likely a loop top source (particle acceleration site). The reconnection or destabilization of the coronal fields is associated with the rising small jets (before first M1.0 flare) and flux rope (during second C8.5 flare). It seems that first reconnection starts in the chromosphere that lead to the formation of two-ribbons, and later reconnection begins in the corona (associated with field lines rotation) resulting the formation of a quasi-circular ribbon. 
The cancellation of the positive flux (P1) (before the trigger of the second C8.5 flare) may contribute to add more flux to the newly formed flux rope to destablize it \citep{sterling2012}. In addition, the continous shear motion between opposite polarity spots most likely helped in the build-up of magnetic energy.  Reconnection below the flux rope may work as a trigger of the flux rope eruption during the C8.5 flare. Periodic reconnection above the flux rope destroyed it and produced untwisting surge like eruption that moved along and across the field lines. This type of behavior is expected in the fan-spine topology \citep{wang2012,kumar2015}. The reconnection of the rising flux tube at the magnetic null-point may be similar to the magnetic breakout reconnection \citep{antiochos1998,sun2013}. 

We agree that the flux rope may emerge below the photosphere, as revealed by the rotation of sunspots in a different event \citep{kumar2013r} as well as in the numerical MHD simulations \citep{fan2009,archontis2008}. But, our observations show the emergence of twisted loops from the opposite polarity spots, and flux rope is formed during magnetic reconnection (in the chromosphere) between two sheared J shaped H$\alpha$ loops. Figure \ref{cartoon}(a,b) displays the schematic cartoon of the scenario, where L1 and L2 join to form a S-shaped flux rope during magnetic reconnection.  Ends of L1 and L2 are anchored in the opposite polarity fields (end of L1, i.e., p in negative and end of L2, i.e., q in positive). They have oppositely directed field lines, which favour the condition of magnetic reconnection between them. Reconnection between the opposite footpoints of L1 and L2 causes the formation of the flux rope in the chromosphere. When q is connected to p, the magnetic pressure is likely to be reduced there. This is because of the connection between p and q in the chromosphere. This allows the expansion of the emerging photospheric flux rope (at q) during the flare, which we observed in the TiO images. Alternatively, formation of the twisted penumbral structure (which we mention flux tube expansion) during the flare may be associated with the back-reaction of magnetic reconnection occurring above it between loops L1 and L2 \citep{hudson2008,wang2013}.

Figure \ref{cartoon}(c) shows the AIA 131 \AA~ reverse color images overlaid by HMI contours of positive (red) and negative (blue) polarities during the second C8.5 flare ($\sim$00:38 and $\sim$00:50 UT). On the basis of observations, we draw the possible magnetic field configuartion (2D) including flux rope reconnection with the overying fields. These panels indicate the plasma flow/surge along the outer spine and associated heating of the arcade loops. In addition, we also noticed the ejections of plasma blobs (upward/downward) along the spine during the second flare.

This scenerio is quite similar to the formation of a coronal flux rope as a result of tether-cutting reconnection \citep{moore2001} between two opposite elbow of the sheared arcade loops. However, our observation shed light on the formation of the chrmospheric twisted flux ropes by magnetic reconnection. In addition, the eruption of the chromospheric flux rope is delayed $\sim$3 hrs after its formation. On the other hand, in tether cutting reconnection, coronal flux rope is erupted in a couple of minutes after its formation. Why did the newly formed flux rope wait for $\sim$3 hrs to erupt? Firstly, the new flux rope formed at a lower height where strapping field may be stronger. Secondly, the twist of the newly formed rope may not be sufficient to lose its equilibrium. HMI data clearly show continuous cancellation/decay of the positive flux after the first M-class flare. This flux might be added to the newly formed flux rope, which may help in the loss of equilibrium of the flux rope. The hard X-ray source above the flux rope supports our interpretation that magnetic reconnection occurred above the rope (similar to magnetic breakout) that destroy/break the flux rope into untwisting surge material. The helical motion of the surge was most likely associated with the chromospheric twisted flux rope.

The appearance of a remote loop heating observed only in the hot AIA channels (94 and 131 \AA) during both flares suggests the link of one footpoint of the heated loop to the reconnection site. This type of remote loop heating is quite common in the magnetic configuration associated with fan-spine topology \citep{sun2013,cheung2015,kumar2015}. In addition, formation of a circular or quasi-circular ribbon is also associated with fan spine topology \citep{masson2009,pariat2010,wang2012,wang2014}. In our case, we observed quasi-circular ribbon and remote loop heating connecting to the flare site. Reconnection at the flare site can cause the acceleration of particles along the remote 94 \AA~ loop, therefore, heating it over $\sim$10 MK temperature.
What exactly triggers the flare in the fan-spine topology configuration that lead to the formation of circular ribbon morphology? Is it slow-mode waves \citep{sych2015} or new emerging flux? Our observations support new flux emergence associated with shear motion as a trigger of the flare, leading to the formation of a twisted flux rope by magnetic reconnection.

Alternative interpretation of this event may be the expansion of the pre-existing emerging flux rope. This interpretation is unlikely because we   see a pre-existing S-shaped flux rope  neither at the photospheric level (in the TiO images) nor at the chromospheric level (H$\alpha$) before the M1.0 flare. We only see separate J shaped strands. Preflare brightening/heating occur exactly at the joining point of the J-shaped loops. Additionally, the TiO images show the rise of the emerging twisting flux-tube (from P2) below the joining point of the J-shaped loops, which is most likely resulted by the reduced magnetic pressure or back reaction of magnetic reconnection on the photosphere as we discussed before (magnetic flux cancellation during the M1.0 flare).

In conclusion, we reported high resolution observations of the formation and eruption a small flux rope. This study highlights the flux rope formation in the chromosphere by magnetic reconnection between two J shaped sheared H$\alpha$ loops during the first M1.0 flare, and its subsequent eruption (like an untwisting surge) during the second C8.5 flare. Observation of plasma inflow in the chromosphere and  magnetic flux cancellation during the first flare support the flux rope formation by magnetic reconnection between two sheared H$\alpha$ loops. Similar events with high-resolution data should be investigated in more detail to understand the dynamics of flux ropes in the chromosphere and their role in triggering the solar flares.

\acknowledgments
 We would like to thank the anonymous referee for his/her constructive comments and suggestions that improved our paper considerably.
SDO is a mission for NASA’s Living With a Star (LWS) program. The SDO data were (partly) provided by the Korean Data Center (KDC) for SDO in cooperation with NASA and SDO/HMI team. RHESSI
is a NASA Small Explorer. BBSO operation is supported by NJIT, US NSF AGS-1250818 and NASA NNX13AG14G, and NST operation is partly supported by the Korea Astronomy and Space Science Institute and Seoul National University. IRIS is a NASA small explorer mission developed and operated by LMSAL with mission operations executed at NASA Ames Research center and major contributions to downlink communications funded by the Norwegian Space Center (NSC, Norway) through an ESA PRODEX contract. HW is supported by US NSF under grants AGS 1348513 and 1408703,  and NASA under grant NNX13AG13G. K.-S. Cho acknowledges support by a grant from the US Air Force Research Laboratory, under agreement number FA 2386-14-1-4078 and by the ``Planetary system research for space exploration" from KASI. This work was supported by the ``Development of Korea Space Weather Center" of KASI and the KASI basic research funds. This work was conducted as part of the effort of NASA's Living with a Star Focused Science Team ``Jets" NASA LWS NNX11AO73G grant.  VYu acknowledges support from NSF AGS-1146896 grant and Korea Astronomy and Space Science Institute.
\bibliographystyle{apj}
\bibliography{reference}

\end{document}